\documentclass[10pt]{article}
\usepackage{amsmath, amssymb, slashed, latexsym, cite, mathrsfs, tensor, tikz-feynman, dsfont, simplewick, xcolor, faktor, float, comment, bbm}
\usepackage[a4paper, left=2.5cm, right=2.5cm, top=3.5cm, bottom=3cm]{geometry}
\usepackage[hidelinks]{hyperref}
\hypersetup{
	colorlinks=true,
	linkcolor=blue,
	citecolor=magenta,
}
\usepackage{booktabs,dcolumn,caption}
\numberwithin{equation}{section}
\makeatletter
\DeclareFontFamily{OMX}{MnSymbolE}{}
\DeclareSymbolFont{MnLargeSymbols}{OMX}{MnSymbolE}{m}{n}
\SetSymbolFont{MnLargeSymbols}{bold}{OMX}{MnSymbolE}{b}{n}
\DeclareFontShape{OMX}{MnSymbolE}{m}{n}{
	<-6>  MnSymbolE5
	<6-7>  MnSymbolE6
	<7-8>  MnSymbolE7
	<8-9>  MnSymbolE8
	<9-10> MnSymbolE9
	<10-12> MnSymbolE10
	<12->   MnSymbolE12
}{}
\DeclareFontShape{OMX}{MnSymbolE}{b}{n}{
	<-6>  MnSymbolE-Bold5
	<6-7>  MnSymbolE-Bold6
	<7-8>  MnSymbolE-Bold7
	<8-9>  MnSymbolE-Bold8
	<9-10> MnSymbolE-Bold9
	<10-12> MnSymbolE-Bold10
	<12->   MnSymbolE-Bold12
}{}
\let\llangle\@undefined
\let\rrangle\@undefined
\DeclareMathDelimiter{\llangle}{\mathopen}%
{MnLargeSymbols}{'164}{MnLargeSymbols}{'164}
\DeclareMathDelimiter{\rrangle}{\mathclose}%
{MnLargeSymbols}{'171}{MnLargeSymbols}{'171}
\makeatother

\newcommand{\im}{\mathrm{i}}
\newcommand{\ep}{\mathrm{e}}
\newcommand{\diff}{\mathrm{d}}
\newcommand{\pa}{\partial}
\newcommand{\tr}{\mathrm{tr}}
\newcommand{\sfrac}[2]{{\textstyle\frac{#1}{#2}}}
\newcommand{\+}{\dagger}
\renewcommand{\=}{\ =\ }
\newcommand{\unity}{\mathbbm{1}}
\newcommand{\und}{\quad\textrm{and}\quad}
\newcommand{\with}{\quad\textrm{with}\quad}
\renewcommand{\>}{\rangle}
\newcommand{\<}{\langle}

\newcommand{\tS}{\ensuremath{\widetilde{S}}}

\newcommand{\bpsi}{{\bar{\psi}}}

\newcommand{\Lag}{\mathcal{L}}
\newcommand{\Dc}{\ensuremath{\mathring{\Delta}}}

\newcommand{\x}{\widetilde{x}}
\newcommand{\Sl}{\stackrel{\leftarrow}{S}}
\newcommand{\Sr}{\stackrel{\rightarrow}{S}}

\newcommand{\drm}{\mathrm{d}}
\newcommand{\rR}{\mathrm{R}}
\newcommand{\cdelta}{\mathring{\delta}}
\newlength{\ucht}
\AtBeginDocument{\settoheight{\ucht}{A}}
\newcommand{\scontour}{\,\begin{tikzpicture}[baseline=1pt]\draw[->] (0,0) -- (1\ucht+1pt,1\ucht+1pt); \end{tikzpicture}\,\,}
\newcommand{\dcontour}{\,\,\begin{tikzpicture}[baseline=1pt]\draw[->] (0,0) -| (1\ucht+1pt,1\ucht+2pt); \end{tikzpicture}\,}
\newcommand{\ucontour}{\,\,\begin{tikzpicture}[baseline=1pt]\draw[->] (0,-1pt) |- (1\ucht+1pt,1\ucht+0pt); \end{tikzpicture}\,\,}

\begin{document}
	\pagenumbering{gobble}
	\sloppy
	\title{\bf\huge \vspace{-25pt}Is the Nicolai map unique?}
	\date{~}
	
	\author{\vspace{-30pt}\phantom{.}\\[12pt]
		{\scshape\Large Olaf Lechtenfeld \ and \ Maximilian Rupprecht}
		\\[24pt]
		Institut f\"ur Theoretische Physik\\ 
		and\\ Riemann Center for Geometry and Physics
		\\[8pt]
		Leibniz Universit\"at Hannover \\ 
		Appelstra{\ss}e 2, 30167 Hannover, Germany
		\\[24pt]
	} 
	
	\clearpage
	\maketitle
	\thispagestyle{empty}
	
	\begin{abstract}
		\noindent\large
		The Nicolai map is a field transformation that relates supersymmetric theories at finite couplings $g$ with the free theory at $g=0$. 
		It is obtained via an ordered exponential of the coupling flow operator integrated from $0$ to $g$. 
		Allowing multiple couplings, we find that the map in general depends on the chosen integration contour in coupling space.
		This induces a large functional freedom in the construction of the Nicolai map, which cancels in all correlator computations.
		Under a certain condition on the coupling flow operator the ambiguity disappears, and the power-series expansion for the map
		collapses to a linear function in the coupling. A special role is played by topological (theta) couplings, 
		which do not affect perturbative correlation functions but also alter the Nicolai map.
		We demonstate that for certain `magical' theta values the uniqueness condition holds, 
		providing an exact map polynomial in the fields and independent of the integration contour.
		This feature is related to critical points of the Nicolai map and the existence of `instantons'.
		As a toy model, we work with $\mathcal{N}=1$ supersymmetric quantum mechanics. For a cubic superpotential and a theta term,
		we explicitly compute the one-, two- and three-point correlation function to one-loop order employing a graphical representation
		of the (inverse) Nicolai map in terms of tree diagrams, confirming the cancellation of theta dependence. 
		Comparison of Nicolai and conventional Feynman perturbation theory nontrivially yields complete agreement,
		but only after adding all (1PI and 1PR) contributions.

	\end{abstract}

	\newpage
	\pagenumbering{arabic}
	
	\section{Introduction, summary and outlook}
	Any off-shell supersymmetric field theory (with coupling parameters~$g$) 
	admits a particular nonlinear and nonlocal field transformation of the bosonic fields~$\phi$, the Nicolai map~$\phi\,\mapsto T_g\phi$. 
	In terms of the (inversely) transformed fields, any correlation function of the interacting theory reduces to a free-field ($g{=}0$) correlator, 
	\begin{equation}
		\bigl\langle X[\phi] \bigr\rangle_g \= \bigl\langle X[T_g^{-1}\phi] \bigr\rangle_0
	\end{equation}
	for any functional~$X[\phi]$, where near the identity $T_g$ can always be inverted perturbatively.
	This formalism was developed by Nicolai, Flume, Dietz and one of the authors in the 1980s~\cite{Nic1,Nic2,Nic3,FL,DL1,DL2,L1}. 
	Recently, the subject has been revisited, extended and freshly illuminated~\cite{ANPP,NP,ALMNPP,AMPP,LR1,MN,LR2,LN,R,M},
	mostly in the context of supersymmetric Yang--Mills theories. 
        Aspects that so far have not received much attention are multiple couplings and topological theta terms in the action.
	These will be elucidated in this work.
	
	Regarding the question in the title, the answer is a resounding `No'! 
	Almost from its inception, it has been clear that the Nicolai map is not unique for gauge theories because it is gauge dependent.
	However, additional freedom was noticed in~\cite{L1} for extended (${\cal N}{=}\,2$) supersymmetry and in four-dimensional super Yang--Mills theory 
	due to the optional $\smallint F\tilde{F}$ term in the action.\footnote{
	    A further possible ambiguity has been noted in~\cite{AMPP} where two distinct maps in six-dimensional super Yang--Mills were found.}
	The first kind of ambiguity has recently been understood to originate from the R-symmetry part of extended supersymmetry,
	as was demonstrated in~\cite{R} for ${\cal N}=\,4$ super Yang--Mills in four dimensions, whose Nicolai map features an~$\mathfrak{su}(4)$ ambiguity in any gauge.
	The second kind of ambiguity arises because a topological term introduces an additional coupling into the theory.
	In fact, this is just the tip of an iceberg!

	We shall argue that, in supersymmetric theories with more than one coupling, there is {\it functional\/} freedom in the Nicolai map.
	We generalize our universal formula~\cite{LR1} in terms of a path-ordered exponential to the multi-coupling setting $g\equiv(g_1,\ldots,g_k)$, 
	\begin{equation} \label{eq:universal_multi}
                T_g[h]\,\phi \= \overrightarrow{\cal P} \exp \Bigl\{-\!\int_{0}^g\!\diff \vec{h}\cdot \vec{R}(h)\Bigr\}\ \phi
		\qquad\textrm{for}\quad \vec{h}=(h_1,\ldots,h_k)\equiv{h} \und \vec{R}_g=(R_{g_1},\ldots,R_{g_k})\ ,
        \end{equation}
	where $R_{g_i}$ generates the coupling flow in the direction of~$g_i$, and
	the line integral runs along a contour (also called~$h$) connecting $0$ and $g$ in coupling space.
	It will be seen that the map in general depends on the chosen contour.
	More concretely, in case of $k$ couplings there is a freedom of $k{-}1$ functions.
	Partial Nicolai maps are defined by flowing only with respect to a subset of couplings while keeping the remaining ones fixed.
	In such a setting, the map will of course still depend on the fixed couplings (in addition to the path for the variable ones).
	Still, it can happen that the Nicolai map is unique, and we shall provide a sufficient condition for it 
	in terms of the coupling flow operator~$\vec{R}_g$. Incidentally, the same condition lets the power-series expansion~\eqref{eq:universal_multi}
	collapse to a linear (in~$g$) expression, i.e.~a polynomial (in the fields) Nicolai map. This can never happen for the inverse map.

	The theta parameter~$\theta$ multiplying a topological term in the action is a special kind of coupling, 
	because it does not affect correlation functions perturbatively.
	However, it turns out that for special `magical' values of~$\theta$ the uniqueness condition is met, and hence
	the map image $T_g[h]\,\phi$ is a unique linear function of~$g$ for any contour~$h$.
	In such cases, the (exact) Nicolai map is related to the existence of `instantons' and carries nonperturbative information 
	about its critical points in field space and spontaneous supersymmetry breaking. 
	For the example of supersymmetric quantum mechanics (SQM), we show that $\theta$ interpolates between two elementary fermion propagators.
	At the magical values $\theta=\pm1$ only one of these is present in the expansion of the Nicolai map, which then can be summed up exactly.
	Unfortunately, outside these special $\theta$~values we cannot yet access the nonperturbative knowledge of~$T_g$,
	but still we may dial~$\theta$ to simplify perturbative computations employing the Nicolai map.

	While for more than one variable coupling the map is path dependent, correlation functions cannot be. 
	We verify this for free massive SQM with a theta term.
	Furthermore, we support our arguments by an explicit one-loop computation of the bosonic one-, two- and three-point correlators 
	in interacting massive SQM, matching the results from Feynman perturbation theory. 
	Reassuringly, the explicit $\theta$~dependence of the inverse Nicolai map always cancels out in the correlation functions.
	It is interesting to compare the Nicolai and Feynman perturbation series. 
	There seem to be separate notions of 1PI diagrams, since the amplitudes agree only when adding all contributions, including the 1PR diagrams.

        A natural next step is explicit computations for more complicated theories, such as sigma models or gauge theories. 
	Perhaps the Nicolai perturbation theory may even contribute to a better understanding of scattering amplitudes in general, 
	as it offers a calculational method distinct from the Feynman perturbation series. 
	Moreover, it will be of great interest to see whether the theta-modified Nicolai map can somehow access 
	nonperturbative (instanton) effects away from the magical $\theta$~values.

	The paper is structured as follows.
	In Section~\ref{sec:nmap} we generalize the Nicolai map to theories with multiple couplings and compare its evaluation for different integration contours. 
	The condition for possible uniqueness of the map is formulated and related to the collapse of the path-ordered exponential.
	Section~\ref{sec:theta} discusses the special role of a topological term, in the context of $\mathcal{N}=\,1$ SQM 
	and highlights the magical $\theta$ values where one can glimpse beyond perturbation theory. 
	We exemplify our findings in Section~\ref{sec:sqm_maps} by performing explicit computations in SQM models with a theta term, 
	firstly for a general superpotential (producing the coupling flow operators), 
	secondly for free massive SQM (giving exact Nicolai maps for different contours),
	and thirdly for an interacting massive SQM (fixing $\theta$ and the mass and setting up the perturbative flow in one coupling).
	For the latter setting we finally provide a diagrammatic perturbation expansion (dubbed `Nicolai rules') 
	and write down a graphical representation for $T_g$ and for $T_g^{-1}$ up to order~$g^4$, which reveals the magic happening at $\theta=\pm1$.
	In Section~\ref{sec:sqm_amplitudes} the one-, two- and three-point correlation function to one-loop order are computed
	for the same model, using Wick-rotated Nicolai rules. 
	The outcomes are indeed independent of~$\theta$ and exactly match the results obtained with conventional (Feynman) perturbation theory, 
	which is performed in Appendix~\ref{app:feyn}. 
	Appendices~\ref{app:straight} and \ref{app:n_proofs} present a more explicit formula for $T_g[h]\,\phi$ for a straight contour in coupling space
	and prove the characteristic (infinitesimal free-action and determinant-matching) properties of the general SQM Nicolai map, respectively.

	\section{The Nicolai map for multiple couplings}\label{sec:nmap}
	Here, we follow the modern definition of the Nicolai map from \cite{LR1}, generalizing the arguments to multiple couplings in a straightforward fashion. 
	We consider a scalar supersymmetric theory and integrate out the fermionic degrees of freedom. The action then takes the form
	\begin{equation}
	S_g[\phi] \= S^b_g[\phi] + \hbar\,S^f_g[\phi]\ ,
	\end{equation}
	where $g=(g_1,\ldots,g_k)$ are coupling constants, which can be seen as local coordinates of some $k$-dimensional coupling space,
	while $S^b_g$ and $S^f_g$ denote the local and nonlocal parts of the action, respectively.
	Expectation values of bosonic observables~$X[\phi]$ are obtained by path integration~\footnote{
	    The vanishing of the vacuum energy in supersymmetric theories properly normalizes $\langle1\rangle_g=1$.}
	\begin{equation} \label{eq:pathintegral}
	\bigl\< X[\phi] \bigr\>_g \= \smash{\int}\!{\cal D}\phi\ \exp\bigl\{ \sfrac{\im}{\hbar} S_g[\phi] \bigr\}\ X[\phi]\ .
	\end{equation}
	Any supersymmetric field theory allows for a
	(generically nonlinear and nonlocal) field transformation, called the Nicolai map
	\begin{equation}
	T_g :\ \phi(x) \ \mapsto\ \phi'(x;g,\phi)
	\end{equation}
	invertible at least as a formal power series in~$g$. Its defining property is
	\begin{equation} \label{eq:globalflow}
	\bigl\< X[\phi] \bigr\>_g \= \bigl\< X[T_g^{-1}\phi] \bigr\>_0
	\qquad\forall\,X \ ,
	\end{equation}
	relating the interacting theory (at couplings~$g$) to the free one (at couplings~$g{=}0$).
	
	\noindent\textbf{Coupling flow operator and Nicolai maps.\ }
	Differentiating~(\ref{eq:globalflow}) with respect to~$g_i$ yields
	\begin{equation} \label{eq:localflow}
	\pa_i \bigl\< X[\phi] \bigr\>_g \= \bigl\< \bigl( \pa_i + R^{(i)}\!(g) \bigr) X[\phi] \bigr\>_g
	\qquad\textrm{for}\quad i=1,\ldots,k \quad\und\quad \pa_i \equiv \tfrac{\pa}{\pa g_i}\ ,
	\end{equation}
	with $k$ functional differential operators~\footnote{
	    We suppress the functional dependence of $R^{(i)}$ on $\phi$ but exhibit its dependence on the collection~$g$ of couplings.}
	\begin{equation}
	R^{(i)}\!(g) \ \equiv\ R_{g_i} \= \int\!\diff x\ \bigl(\pa_i T_g^{-1} \circ T_g \bigr) \phi(x)\,\frac{\delta}{\delta\phi(x)}
	\ =:\ \int\!\diff x\ K_i[\phi;\;x]\,\frac{\delta}{\delta\phi(x)}\ .
	\end{equation}
	The map~$T_g$ may be found from the relations
	\begin{equation} \label{eq:kernel}
	\bigl( \pa_i + R^{(i)}\!(g)\bigr)\,T_g \phi \= 0\ ,
	\end{equation}
	which immediately follow from~\eqref{eq:globalflow} for $X[\phi]=T_g\phi$. The solution is the path-ordered exponential
	\begin{equation}\label{eq:po_exp}
	T_g[h]\,\phi \= \overrightarrow{\cal P} \exp \Bigl\{-\!\int_{0}^1\!\diff s\ h'_i(s)\ R^{(i)}\!\bigl(h(s)\bigr)\Bigr\}\ \phi 
	\qquad\textrm{for}\quad h_i(0)=0 \und h_i(1)=g_i\ ,
	\end{equation}
	generalized to multiple variables $g=\{g_i\}$ and depending on a path $h(s)=\bigl(h_1(s),\ldots,h_k(s)\bigr)$ in coupling space. 
	It can also be formally inverted in a straightforward fashion, by reversing the path-ordering and the sign in the exponential. 
	We may integrate over any path from $0$ to $g$. 
	It is important to understand the path dependence of the Nicolai map indicated in the notation. 
	For correlation functions, we surely conclude
	\begin{equation} \label{eq:flatness}
	\partial_i\partial_j\langle X\rangle_g\=\partial_j\partial_i\langle X\rangle_g \qquad\Rightarrow\qquad
	\bigl\langle \partial_i \bigl(R^{(j)}X\bigr) -\partial_j \bigl(R^{(i)}X\bigr) +[R^{(i)},\,R^{(j)}]\,X \bigr\rangle_g\=0\ ,
	\end{equation}
	which for $X=\unity$ yields
	\begin{equation} \label{eq:flatness_cond}
	\bigl\langle \pa_i R^{(j)} - \pa_j R^{(i)} + [R^{(i)},\,R^{(j)}] \bigr\rangle_g \= 0\ .
	\end{equation}
	This may be interpreted as a weak flatness property for a `flow one-form' field
	\begin{equation}
	R(g)\=\sum_{i=1}^k \drm g_i\,R^{(i)}\!(g)\ .
	\end{equation}
	By a generalized Stokes theorem this implies that the averaged holonomy of $R(g)$ is trivial. 
	In other words, expectation values such as \eqref{eq:globalflow} do not depend on the integration contour~$h$. 
	However, this does not imply that the Nicolai map itself is path-independent, on the contrary! 
	We will see explicitly in subsection~\ref{subsec:massive} that this is not the case. 
	Hence, for theories with multiple couplings, we encounter a large ambiguity for the Nicolai map, parametrized by a curve in coupling space.
	
	This ambiguity can be used to define `partial Nicolai maps'. The composition 
	\begin{equation}
	T_{g\tilde{g}}[\tilde{h}^{-1}\circ h]\;\phi \ :=\ T_{g}[h]\; T^{-1}_{\tilde{g}}[\tilde{h}^{-1}]\;\phi 
	\qquad\textrm{for}\quad \tilde{h}_i(1)=\tilde{g}_i \und  h_i(1)=g_i
	\end{equation}
	allows us to connect any two values in coupling space and, in particular, to flow inside a coupling {\it subspace\/} only.
	Rather than using the above composition, for such a situation it is preferable to employ adapted coordinates in coupling space
	and exponentiate only the flow operators inside the subspace, while keeping fixed the external couplings. 
	The resulting partial Nicolai map will turn on just the couplings in the subspace (along some path inside it) 
	but will still depend on the fixed values of the external couplings.
	For the full Nicolai map (which controls all couplings), two special cases will be investigated more closely in the following,
	\begin{equation}
	\begin{aligned}
	\textrm{sequential flow:} \qquad &(0,0,0,\ldots,0)\ \mapsto\ (g_1,0,0,\ldots,0)\ \mapsto\ (g_1,g_2,0,\ldots,0)\ \mapsto\ \ldots\ \mapsto\  (g_1,g_2,g_3,\ldots,g_k) \ ,\\
	\textrm{straight flow:}   \qquad &h_i(s) = s\,g_i \qquad\textrm{for}\quad s\in[0,1] \ .
	\end{aligned}
	\end{equation}
	In the later parts of this work, we mostly consider one-variable Nicolai maps in the presence of two or three couplings. 
	
	For completeness, we note the characteristic properties which follow from the defining relation \eqref{eq:globalflow} of the Nicolai map. 
	In terms of path integrals and collecting powers of $\hbar$, one finds (for any path~$h$!) that
	\begin{equation}
		S^{\mathrm{b}}_0[T_g\phi] \= S^{\mathrm{b}}_g[\phi] \quad\und\quad
		S^{\mathrm{f}}_0[T_g\phi] -\im\,\tr\ln\sfrac{\delta T_g\phi}{\delta\phi} \= S^{\mathrm{f}}_g[\phi]\ ,
	\end{equation}
	the `free-action' and `determinant-matching' property, respectively. 
	The equivalent infinitesimal conditions for the coupling flow operator follow from \eqref{eq:localflow}~\cite{L2}, 
	\begin{equation}\label{eq:cf_cond12}
		\bigl(\partial_i+R^{(i)}\!(g)\bigr)\,S^{\mathrm{b}}_g[\phi]\=0 \quad\und\quad 
		\bigl(\partial_i+R^{(i)}\!(g)\bigr)\,S^{\mathrm{f}}_g[\phi]\=\im\int\!\diff x\ \frac{\delta K_i[\phi;\;x]}{\delta \phi(x)}\ .
	\end{equation}
	
	\noindent\textbf{Explicit integration contours.\ }
	Let us expand the universal formula \eqref{eq:po_exp} in obvious shorthand notation,
	\begin{equation} \label{eq:po_exp_exp}
	\begin{aligned}
	T_g[h]\,\phi& \= \overrightarrow{\cal P} \exp \Bigl\{-\!\int_{0}^1\!\diff s\ \vec{h}'(s)\cdot \vec{R}\bigl(h(s)\bigr)\Bigr\}\ \phi\\
	& \= \sum_{n=0}^{\infty}(-)^n\int_{0}^{1}\drm s_n\int_0^{s_n}\drm s_{n-1}\cdots\int_0^{s_2}\drm s_1\ 
	\Bigl[\vec{h}'(s_n)\cdot \vec{R}\bigl(h(s_n)\bigr)\Bigr]\cdots\Bigl[\vec{h}'(s_1)\cdot \vec{R}\bigl(h(s_1)\bigr)\Bigr]\ \phi\ .
	\end{aligned}
	\end{equation}
	For simplicity, we temporarily restrict to two couplings $g_1$ and $g_2$, but provide a formula for a general straight flow
	in Appendix \ref{app:straight}. 

	Integrating along the straight contour
	$$
	\begin{tikzpicture}[decoration={markings,
		mark=at position 1cm   with {\arrow[line width=1pt]{stealth}},
		mark=at position 2cm with {\arrow[line width=1pt]{stealth}},
	}]
	\draw[thick, ->] (0,0) -- (2.5,0) coordinate (xaxis);
	
	\draw[thick, ->] (0,0) -- (0,2.5) coordinate (yaxis);
	
	\coordinate (1) at (2,2);
	
	\node[above] at (xaxis) {$h_1$};
	\node[right]  at (yaxis) {$h_2$};
	
	\path[draw,blue, line width=0.8pt, postaction=decorate] 
	(0,0)
	--  (1)  node[above right, black] {$g$};
	
	\fill (1) circle[radius=2pt];
	\end{tikzpicture}$$
	with
	\begin{equation}
	h_1(s)\=s\,g_1\und h_2(s)\=s\,g_2\ ,
	\end{equation}
	and expanding the operators
	\begin{equation}\label{eq:R_exp}
	\begin{aligned}
	&R^{(1)}(g) \= \sum_{k=1}^\infty\sum_{l=0}^\infty g_1^{k-1}g_2^{l} \rR^{(1)}_{k,l} \= 
	\rR^{(1)}_{1,0} + g_1\,\rR^{(1)}_{2,0} + g_2\,\rR^{(1)}_{1,1} + g_1g_2 \rR^{(1)}_{2,1} + \ldots\ ,\\
	&R^{(2)}(g) \= \sum_{k=0}^\infty\sum_{l=1}^\infty g_1^{k}g_2^{l-1} \rR^{(2)}_{k,l} \= 
	\rR^{(2)}_{0,1} + g_2\,\rR^{(2)}_{0,2} + g_1\,\rR^{(2)}_{1,1} + g_1g_2 \rR^{(2)}_{1,2} + \ldots\ ,
	\end{aligned}
	\end{equation}
	the final formula for the two couplings and in case of a straight flow can be written as (see Appendix~\ref{app:straight})
	\begin{equation}
		T_g[\scontour]\;\phi
		\= \sum_{n=0}^{\infty}\ \sum_{i_1,\ldots, i_n=1}^2 \sum_{\alpha}\ c_\alpha\ g^\alpha\ 
		\rR^{(i_n)}_{\alpha^n}\cdots \rR^{(i_1)}_{\alpha^1}\ \phi\ ,
	\end{equation}
	where $\alpha=(\alpha^1,\ldots,\alpha^n)$ is a multi-index of $n$ pairs ($i=1,\ldots,n$)
	\begin{equation}
		\alpha^i\=\bigl((\alpha^i)_1,(\alpha^i)_2\bigr)
	\end{equation}
	that takes values
	\begin{equation}
		(\alpha^p)_{i_q}\ \geq\ 0\quad \text{for}\quad p\ \neq\ q\ ,\qquad\qquad (\alpha^p)_{i_p}\ \geq\ 1\ ,
	\end{equation}
	and
	\begin{equation}
		g^{\alpha}\=g^{\alpha^1}\cdots\ g^{\alpha^n}\=g_1^{\sum_{l=1}^{n}(\alpha^l)_1} g_2^{\sum_{l=1}^{n}(\alpha^l)_2}\ .
	\end{equation}
	One finds for the first few terms
	\begin{equation}\label{eq:nmap_sym}
	\begin{aligned}
	T_g[\scontour]\,\phi\=\phi\ -\ &g_1\rR^{(1)}_{1,0}\;\phi\ -\ g_2\rR^{(2)}_{0,1}\;\phi\ 
	-\ \sfrac12 g_1^2\bigl(\rR^{(1)}_{2,0}-\rR^{(1)}_{1,0}\rR^{(1)}_{1,0}\bigr)\;\phi\ 
	-\ \sfrac12 g_2^2\bigl(\rR^{(2)}_{0,2}-\rR^{(2)}_{0,1}\rR^{(2)}_{0,1}\bigr)\;\phi\\ 
	-\ &\sfrac12 g_1g_2\bigl(\rR^{(1)}_{1,1}+\rR^{(2)}_{1,1}-\rR^{(1)}_{1,0}\rR^{(2)}_{0,1}-\rR^{(2)}_{0,1}\rR^{(1)}_{1,0}\bigr)\;\phi\ +\  \ldots\ .
	\end{aligned}
	\end{equation}
	
	Consider instead the sequential contour
	$$
	\begin{tikzpicture}[decoration={markings,
		mark=at position 1cm   with {\arrow[line width=1pt]{stealth}},
		mark=at position 3cm with {\arrow[line width=1pt]{stealth}},
	}]
	\draw[thick, ->] (0,0) -- (2.5,0) coordinate (xaxis);
	
	\draw[thick, ->] (0,0) -- (0,2.5) coordinate (yaxis);
	
	\coordinate (1) at (2,0);
	\coordinate (2) at (2,2);
	
	\node[above] at (xaxis) {$h_1$};
	\node[right]  at (yaxis) {$h_2$};
	
	\path[draw,blue, line width=0.8pt, postaction=decorate] 
	(0,0)
	--  (1) -- (2)  node[above right, black] {$g$};
	
	\fill (2) circle[radius=2pt];
	\end{tikzpicture}$$
	one can again write down the first few terms of the expansion and carry out the integrals.
	Here we do not need the $g_2$-dependence of $R^{(1)}$, as can be expected intuitively from the integration contour. 
	Expanding the operators as in \eqref{eq:R_exp}, we obtain
	\begin{equation}\label{eq:nmap_asym}
	\begin{aligned}
	T_g[\dcontour]\,\phi\=
	\phi\ -\ &g_1\rR^{(1)}_{1,0}\;\phi\ -\ g_2\rR^{(2)}_{0,1}\;\phi\ 
	-\ \sfrac12 g_1^2\bigl(\rR^{(1)}_{2,0}-\rR^{(1)}_{1,0}\rR^{(1)}_{1,0}\bigr)\;\phi\ 
	-\ \sfrac12 g_2^2\bigl(\rR^{(2)}_{0,2}-\rR^{(2)}_{0,1}\rR^{(2)}_{0,1}\bigr)\;\phi\\ 
	-\ & g_1g_2\bigl(\rR^{(2)}_{1,1}-\rR^{(2)}_{0,1}\rR^{(1)}_{1,0}\bigr)\;\phi\ +\  \ldots\ .
	\end{aligned}
	\end{equation}
	Clearly, the two maps \eqref{eq:nmap_sym} and \eqref{eq:nmap_asym} start to differ at order $g_1g_2$. 
	The weak flatness \eqref{eq:flatness} however implies that
	\begin{equation} \label{eq:weakg1g2}
	\bigl\langle (\rR^{(1)}_{1,0}\rR^{(2)}_{0,1}-\rR^{(2)}_{0,1}\rR^{(1)}_{1,0})\;\phi \bigr\rangle_g 
	\= \bigl\langle (\rR^{(1)}_{1,1}-\rR^{(2)}_{1,1})\;\phi\bigr\rangle_g\ ,
	\end{equation}
	so that the difference does not matter inside correlation functions.
	
	\noindent\textbf{A condition for contour independence.\ }
        Is it possible that the integration contour ambiguity of the Nicolai map disappears, rendering the map unique after all?
        Let us look into this again for the case of two couplings $g_1$ and $g_2$ and compare the two different sequential contours.
        From the structure of the leading path dependence~\eqref{eq:weakg1g2} and higher-order analogs,
        one sees that a sufficient condition is
        \begin{equation}\label{eq:collapse_cond}
        \begin{aligned}
        \rR^{(1)}_{k,l}\,\rR^{(1)}_{1,0}\,\phi &\= \rR^{(1)}_{k+1,l}\,\phi \quad\und\quad
        \rR^{(1)}_{k,l}\,\rR^{(2)}_{0,1}\,\phi \= \rR^{(1)}_{k,l+1}\,\phi \ ,\\
        \rR^{(2)}_{k,l}\,\rR^{(1)}_{1,0}\,\phi &\= \rR^{(2)}_{k+1,l}\,\phi \quad\und\quad
        \rR^{(2)}_{k,l}\,\rR^{(2)}_{0,1}\,\phi \= \rR^{(2)}_{k,l+1}\,\phi
        \end{aligned}
        \end{equation}
        for all possible values of $k$ and $l$. If these relations are met, however, the map becomes unique, because
		all terms in the expansions \eqref{eq:nmap_sym} and~\eqref{eq:nmap_asym} beyond the linear part will vanish!
        To see this, pair up all terms whose $\rR$-strings end like the two sides in the above relations but are equal otherwise.
        The two multiple $s$-integrals for any pair turn out to be equal, and so the factor $(-)^n$ in~\eqref{eq:po_exp_exp}
        lets the two terms in any pair cancel each other. In this way,
        all contributions except for those from a single action of $\rR^{(1)}_{1,0}$ or $\rR^{(2)}_{0,1}$ disappear,
        and one is left with
        \begin{equation}\label{eq:collapse_one_coupling}
        T_g\,\phi \= \phi\ -\ g_1\rR^{(1)}_{1,0}\;\phi\ -\ g_2\rR^{(2)}_{0,1}\;\phi\ ,
        \end{equation}
        which constitutes an exact Nicolai map linear in the couplings!
        Usually $R^{(i)}(g{=}0)$ is polynomial in the fields~$\phi$, and then this map is also just polynomial.
        The generalization to more than two couplings is obvious.

        We learn that the criterion for uniqueness (at more than one coupling) is also sufficient for polynomiality of the map.
        For a single coupling, the Nicolai map is always unique, but the analog of the expansions~\eqref{eq:R_exp},
	\begin{equation} \label{eq:Rg_exp}
	R_g(g) \= \sum_{k=0}^\infty g^k\,\rR_{k+1} \= \rR_1 + g\;\rR_2 + g^2\,\rR_3 + \ldots\ ,
	\end{equation}
	and of the relations~\eqref{eq:collapse_cond},
        \begin{equation}\label{eq:uniqueness_cond}
        \rR_k\,\rR_1\;\phi \= \rR_{k+1}\;\phi \quad\textrm{for}\quad k\ge1\ ,
        \end{equation}
        still truncates the power series \eqref{eq:po_exp_exp} to its linear part in~$g$,
	\begin{equation}
	T_g\;\phi \= \phi\ -\ g\,\rR_1\;\phi\ .
	\end{equation}
        Note that this can never happen for the inverse map, because all contributions come with the same sign.
        We will demonstrate this mechanism explicitly in Section~4 for the magical theta values $\theta{=}{\pm}1$ in SQM.

        The uniqueness condition~\eqref{eq:uniqueness_cond} has a graphical interpretation:
        Every action of the coupling-flow operator grafts a linear tree, i.e.~one with an unbranched fermion line.
        Hence, $R^2\phi$ produces all kinds of trees with one or two fermion branches.
        Truncating this to the unbranched (linear) trees and remembering~\eqref{eq:Rg_exp}
        we may rewrite \eqref{eq:uniqueness_cond} as
        \begin{equation}
        R_g^2(g)\;\phi\big|_{\textrm{linear}} \= \partial_g R_g(g)\;\phi\ .
        \end{equation}

	\section{The special role of theta}\label{sec:theta}
	Sometimes the action may be amended by a so-called topological term, with its own coupling $\theta$, 
	which captures topologically nontrivial field configurations and becomes relevant non-perturbatively.
	Such a `theta term' in the Lagrangian takes the form of a total derivative and therefore contributes
	to the action only when the field has nontrivial structure `at infinity'. In the Euclidean path integral,
	this happens when the field makes excursions to critical points of the action other than the perturbative vacuum.
	Such `instanton' configurations are suppressed by a factor of $\exp(-|k|\,c/g^2)$, 
	where $c$ is a fixed positive coefficient and $k\in\mathbb{Z}$ characterizes the field topology. 
	A theta term in the action will modify the coefficient~$c$ in the weight factor above.
	Perturbation theory around a stable vacuum can only access the $k{=}0$ sector,
	hence will not produce any $\theta$~dependence in quantum correlation functions.
	Prominent examples are (super) Yang--Mills theory and (supersymmetric) quantum mechanics.
	
	In this section we shall investigate the special role of a theta term for the Nicolai map.
	As any coupling parameter does, $\theta$ comes with its own theta flow operator~$R_\theta$ 
	but also modifies the other coupling flow operators~$R_g$. 
	We will show that the theta flow is trivial in perturbation theory, 
	thus perturbative correlators are $\theta$-independent. 
	However, the other coupling flows, and thus the Nicolai map, turn out to depend on~$\theta$
	in an interesting fashion. Employing the inverse Nicolai map in the computation of
	$N$-point functions must then conspire to cancel all $\theta$-dependent contributions.
	
	For definiteness, we will elucidate these properties in the example of
	one-dimensional supersymmetric quantum mechanics, whose action is given by
	\begin{equation}
		S[x;g,\theta] \= \int\!\diff t\ \Bigl\{ \tfrac12 \dot x^2 - \tfrac12 V'(x)^2 + 
		\bar\psi \bigl[ \im\sfrac{\diff}{\diff t} - V''(x) \bigr] \psi + \im\theta\tfrac{\diff}{\diff t}V(x) \Bigr\}
	\end{equation}
	with $\tfrac{\diff}{\diff t}V(x)=\dot x\,V'(x)$ for the $(0{+}1)$-dimensional `field' $x(t)$, 
	where one or more couplings~$g$ are hiding in the superpotential~$V(x)$. 
	For a review, see~\cite{CKS}. Early important papers are~\cite{Witten,vanHolten,SH,CF}.
	The simplest nontrivial superpotential is given by
	\begin{equation}
		V'(x) \= m\,x + g\,x^2 \= g\bigl(x+\tfrac{m}{2g}\bigr)^2 - \tfrac{m^2}{4g} 
	\end{equation}
	with two couplings $g$ and~$m$.
	We note that this action has two involution symmetries,
	\begin{equation} \label{eq:involutions}
		(x,\psi,\bar\psi;g,m,\theta)\ \mapsto\ (-x,\psi,\bar\psi;-g,m,\theta) \quad\und\quad
		(x,\psi,\bar\psi;g,m,\theta)\ \mapsto\ (-x,\bar\psi,\psi;g,-m,-\theta)\ .
	\end{equation}
	The corresponding bosonic potential $\tfrac12 {V'}^2$ is of the double-well type,
	with minima at $x{=}0$ and $x{=}-m/g$. This is the simplest instance of non-perturbative
	supersymmetry breaking~\cite{Witten}. The ground state remains doubly degenerate,
	and its energy is lifted from zero by instanton-antiinstanton contributions and thus
	of the order $\exp(-2\,c/g^2)$ with $c{=}m^3/6$ for $\theta{=}0$.
	
	As explained in the previous section, we may flow from vanishing coupling to a point
	$(g,m,\theta)$ in coupling space along any path we like. Let us choose
	\begin{equation}
		(0,0,0)\ \mapsto\ (0,0,\theta)\ \mapsto\ (g,m,\theta)\ ,
	\end{equation}
	so that $\theta$ is turned on in the free (massless) theory first. However, since
	\begin{equation}
		\partial_\theta S[x;g,m,\theta] \= \im\bigl[ \tfrac12 mx^2 + \tfrac13 gx^3 \bigr]_{t=-\infty}^{t=+\infty}
	\end{equation}
	vanishes for $g{=}m{=}0$, the theta flow operator $R_\theta$ vanishes there, 
	and hence the finite Nicolai transformation remains the identity until $(0,0,\theta)$.
	We note that this feature does not occur for standard couplings, which may be turned on 
	in the absence of other parameters, i.e.~in the free theory.
	However, this does not yet imply that the further flow to $(g,m,\theta)$ remains
	$\theta$ independent. On the contrary, via $R_{g,m}(g,m,\theta)$ it will introduce 
	a definite $\theta$~dependence into the final Nicolai map.
	
	We can actually be slightly more explicit, because turning on $m$ brings us from the free
	massless to a free massive theory, where everything is still computable analytically.
	In the following section, before turning to the interacting model, we will show that moving along
	\begin{equation}
		(0,0,0)\ \mapsto\ (0,m,\theta)  
	\end{equation}
	in different ways creates closed-form expressions for the $m$ and $\theta$ dependence of all quantities
	and establishes $\theta$-independence of all (free massive) correlators.
	Verifying the latter property for the interacting theory is nontrivial but nevertheless true
	perturbatively, as we shall then demonstrate.
	
	Let us return to a general superpotential $V(x)$ depending on one or more couplings~$g$
	and hold the special parameter~$\theta$ at a fixed value (no $\theta$~flow).
	Since any flow operator $R^{(i)}(g,\theta)$ is linear in~$\theta$, the Nicolai map
	$T_g[h,\theta]$ and its inverse (for a given contour~$h$ in $g$-space) are power series in~$g$, 
	where the order $g^n$ term (in a multi-index sense) is a polynomial of degree~$n$ in~$\theta$.
	When $\theta{=}\pm1$ (and only for these two values), some magic happens:
	For any choice $h$ of integration contour, the series for $T_g[h,\pm1]$ truncates to become linear in~$V'$!
	Concretely, one finds that~\footnote{The full justification will be presented in Section~\ref{sec:sqm_maps} together with a proof of \eqref{eq:collapse_one_coupling}.}
	\begin{equation} \label{eq:polyTdot}
		\im\tfrac{\diff}{\diff t} T_g[h,\pm1]\;x(t) \= 
		T_g[h,\pm1]\;\im\dot{x}(t) \=
		\im\dot x(t)\ \mp\ V'\bigl(x(t)\bigr)\ .
	\end{equation}
	It may be disturbing that this map is complex, but it still leads to real correlation functions.
	
	The magical values $\theta=\pm1$ allow for a direct connection to instantons and 
	tell us about the range of invertibility of the Nicolai map.
	The solutions $\bar x(t)$ to
	\begin{equation}
		T_g \bar x(t)\=\textrm{constant} \qquad\Rightarrow\qquad 
		\im \dot{\bar x}(t)\ \mp\ V'\bigl(\bar x(t)\bigr) \= 0
	\end{equation}
	represent instantons or antiinstantons (depending on the sign) interpolating between
	neighboring zeros of $V'$, after a Wick rotation to imaginary time.
	It is well known that the velocity~$\dot{\bar x}$ is a zero mode of the fermion fluctuation operator
	about the instanton configuration, the Goldstone mode of broken time translation invariance:
	\begin{equation}
		\bigl[ \im\sfrac{\diff}{\diff t} - V''(\bar x) \bigr]\,\dot{\bar x} \= 0\ .
	\end{equation}
	But for $\theta=+1$, the fluctuation operator agrees with the Jacobian of the Nicolai map,\footnote{
	    up to a trivial $\im\sfrac{\diff}{\diff t}$ factor. 
	    For $\theta=-1$, we can flip the sign of $V''$ by applying the second involution in \eqref{eq:involutions}.}
	\begin{equation}
		\im\sfrac{\diff}{\diff t} \frac{\delta T_g x(t)}{\delta x(t')} \= \bigl[ \im\sfrac{\diff}{\diff t} - V''(x(t)) \bigr]\,\delta(t{-}t')\ ,
	\end{equation}
	and thus this Jacobian develops a zero eigenvalue for $x=\bar x$.
	In other words, the trivial invertibility of the Nicolai map for configurations~$x$
	near the vacuum $x{\equiv}0$ breaks down (for $\theta{=}\pm1$) when $x$ grows to 
	an instanton~$\bar x$ if not earlier.
	This is in tune with the interpretation of the Witten index as the mapping degree of
	the Nicolai map, which is obtained from counting its critical points (with sign).
	It also shows that there is nonperturbative information in the full Nicolai map,
	even though its extraction requires summing up its series, 
	which we currently can manage only for the magical $\theta$~values.

	\section{Nicolai maps for supersymmetric quantum mechanics}\label{sec:sqm_maps}

	In Section~\ref{sec:nmap}, we established how the Nicolai map (with multiple couplings) 
	can be obtained perturbatively through the coupling flow operators. 
	We can construct the latter canonically \cite{L1,LR1} from the off-shell supersymmetric action. 
	In the following, we first introduce $D$-dimensional supersymmetric quantum mechanics (SQM) 
	for a generic superpotential in \ref{subsec:arbitrary}, deducing from it the coupling flow operators 
	for $\theta$ and regular couplings $g$. 
	Next, we specialize to a free massive theory (couplings $m$ and $\theta$) in~\ref{subsec:massive}, 
	where we can find explicit expressions for the flow operators $R_m$ and $R_\theta$ 
	as well as the corresponding (inverse) Nicolai maps in frequency space. 
	As a more complex model, we then consider a massive interacting theory
	with a cubic term $\sim g\,x^3$ in the superpotential in~\ref{subsec:interacting}. 
	This theory is controlled by the parameters $g$,$m$ and~$\theta$. 
	To avoid infrared divergences in the computation of correlators later on, 
	we find it most convenient to work with the one-variable inverse Nicolai map $T^{-1}_g$ 
	at fixed mass $m\neq 0$ and $\theta$. Finally, we express this map in a compact graphical notation, 
	which we christian `Nicolai rules', in~\ref{subsec:Nicolai_rules}.

	\subsection{General superpotential}\label{subsec:arbitrary}
	\noindent\textbf{Construction of the SQM action.\ }
	The off-shell action of $D$-dimensional SQM theory with a topological theta term reads
	\begin{equation}\label{eq:off_shell_action}
		\mathring{S}\=
		\smash{\int} \!\diff t\ \Bigl\{\sfrac{1}{2}\dot{x}_i^2+\sfrac12 A_i^2- A_iV_i(x)+\bpsi_i\bigl[\im\sfrac{\diff}{\diff t}\delta_{ij}-V_{ij}(x)\bigr]\psi_j
		+\im\theta \sfrac{\diff}{\diff t}V\bigl(x(t)\bigr)\Bigr\}\ ,
	\end{equation}
	with bosonic variables $x_i(t)$, fermionic variables $\psi_i(t)$ and $\bpsi_i(t)$ as well as bosonic auxiliary variables $A_i(t)$, for $i=1,\ldots,D$.
	The theory is determined by a superpotential $V(x)$, with derivatives $V_i=\partial V/\partial x_i$ 
	and $V_{ij}=\partial^2 V/(\partial x_i\,\partial x_j)$. 
	The off-shell supervariations can be written in terms of two fermionic operators $\cdelta$ and $\bar{\cdelta}$,
	\begin{equation}\label{eq:off_shell_variations}
	\begin{aligned}
	&\cdelta x_i\=\psi_i\ ,\qquad \cdelta\psi_i\=0\ ,\qquad \cdelta\bpsi_i\=\im\dot{x}_i- A_i\ ,\qquad\cdelta A_i\=\im\dot{\psi}_i\ ,\\
	&\bar{\cdelta} x_i\=\bpsi_i\ ,\qquad \bar{\cdelta}\psi_i\=\im\dot{x}_i+A_i\ ,\qquad \bar{\cdelta}\bpsi_i\=0\ ,\qquad\bar{\cdelta} A_i\=-\im\dot{\bpsi}_i\ .
	\end{aligned}
	\end{equation}
	The Lagrangian in~\eqref{eq:off_shell_action} may be obtained through a superspace formalism as the last component of a superfield,
	and thus is a supervariation of the penultimate components of this superfield. The off-shell action integral may therefore be written as
	\begin{equation}
        \mathring{S}\ \=\ \sfrac12\bigl(\cdelta\bar{\Dc}+\bar{\cdelta}\Dc\bigr)\ +\ \sfrac{\theta}{2}\bigl(\cdelta\bar{\Dc}-\bar{\cdelta}\Dc\bigr)
	\ \=\ \sfrac{1+\theta}{2}\,\cdelta\bar{\Dc}\ +\ \sfrac{1-\theta}{2}\,\bar{\cdelta}\Dc\ ,
        \end{equation}
	with the integrated superfield components
	\begin{equation}
	\Dc\=\smash{\int} \!\diff t\ \bigl\{ \sfrac12 (A_i-\im\dot{x}_i)- V_i(x)\bigr\}\, \psi_i \quad\und\quad
	\bar{\Dc}\=\smash{\int} \!\diff t\ \bigl\{ \sfrac12(-A_i-\im\dot{x}_i)+ V_i(x)\bigr\}\, \bpsi_i\ .
	\end{equation}
	After eliminating the auxiliary field via its equation of motion $A_i=V_i$, the on-shell action reads
	\begin{equation}\label{eq:full_action}
		S \=\smash{\int} \!\diff t\ \Bigl\{ \sfrac{1}{2}\dot{x}_i^2-\sfrac12 V_i(x)^2
		+\bpsi_i\bigl[\im\sfrac{\diff}{\diff t}\delta_{ij}-V_{ij}(x)\bigr]\psi_j+\im\theta \sfrac{\diff}{\diff t}V\bigl(x(t)\bigr)\Bigr\}\ .
	\end{equation}
	Contrary to the full action,\footnote{
            The correct on-shell action \eqref{eq:full_action} requires taking the off-shell supervariations $\cdelta$ and $\bar{\cdelta}$                  
            of the off-shell integrals $\Dc$ and $\bar{\Dc}$, and {\it then\/} imposing $A_i=V_i$. The other way around yields a wrong coefficient for the fermion term.}
	the $\theta$ term may be generated on-shell,
	\begin{equation} \label{eq:theta_derivative}
	    \partial_\theta S \= \sfrac12\bigl(\cdelta\bar{\Dc}-\bar{\cdelta}\Dc\bigr) 
	    \= \sfrac12 \bigl( \delta\bar\Delta^\beta - \bar\delta\Delta^\beta \bigr)
	    \= \smash{\int} \!\diff t\ V_i(x)\ \dot x_i \= \im\,V\bigl(x(t)\bigr)\big|^{+\infty}_{-\infty}
	\end{equation}
	with the on-shell variations $\delta$ and $\bar{\delta}$ 
	(obtained from $\cdelta$ and $\bar{\cdelta}$ by setting $A_i=V_i$ in \eqref{eq:off_shell_variations}) as well as
	\begin{equation} \label{eq:thet_Delta_generation}
        \Delta^\beta\=\smash{\int} \!\diff t\ \bigl\{ (\beta{-}1)\im\dot{x}_i -\beta V_i(x)\bigr\}\,\psi_i \quad\und\quad
        \bar{\Delta}^\beta\=\smash{\int} \!\diff t\ \bigl\{ (\beta{-}1)\im\dot{x}_i +\beta V_i(x)\bigr\}\,\bpsi_i
	\end{equation}
	including an ambiguity $\beta\in\mathbb{R}$. 
	We note that $\Delta^{\beta=\frac12}=\Dc\big|_{A_i=V_i}$ and that $\Delta^{\beta=1}\big|_{V=0}=0$.
	
	\noindent\textbf{Construction of the $g$- and $\theta$-flow operators.\ }
	We consider an arbitrary superpotential $V$ that depends on a coupling $g$.\footnote{
	    If $V$ depends on multiple couplings, the following discussion can be applied to each of the couplings separately.} 
	The $g$-derivative of the action can, just like the action itself, be written as a supervariation, but we may go on-shell now:
	\begin{equation}
	\partial_g S\= (\partial_g \mathring{S})\big|_{A_i=V_i} \= \sfrac{1+\theta}{2}\,\delta\bar{\Delta}_g\ +\ \sfrac{1-\theta}{2}\,\bar{\delta}\Delta_g
	\end{equation}
	with the simpler integrals
	\begin{equation}
	\Delta_g\ \equiv\ \partial_g\Dc\=-\smash{\int}\!\diff t\ \bigl\{\psi_i\ \partial_g V_i\bigr\} \quad\und\quad 
	\bar{\Delta}_g\ \equiv\ \partial_g\bar{\Dc}\=\smash{\int}\!\diff t\ \bigl\{\bpsi_i\  \partial_g V_i\bigr\}\ .
	\end{equation}
	Employing the supersymmetric Ward identity as usual \cite{L1}, this leads to the $g$-flow operator
	\begin{equation}
	\begin{aligned}
	R_g(g,\theta)&\=
	\sfrac{1+\theta}{2}\,\im\bcontraction{}{\bar{\Delta}}{_g\ }{\delta} \bar{\Delta}_g\ \delta\ +\ 
	\sfrac{1-\theta}{2}\,\im \bcontraction{}{\Delta}{_g\ }{\bar{\delta}} \Delta_g\ \bar{\delta}\\
	&\= \int\!\diff t\ \diff t'\ \sfrac{\im}{2}(\partial_g V_i)(t)\ \bigl\{ 
	(1{+}\theta)\;\bcontraction{}{\bpsi}{_i(t)\,}{\psi} \bpsi_i(t)\,\psi_j(t')\;-\;
	(1{-}\theta)\;\bcontraction{}{\psi}{_i(t)\,}{\bpsi} \psi_i(t)\,\bpsi_j(t') \bigr\}\ 
	\sfrac{\delta}{\delta x_j(t')}\ ,
	\end{aligned}
	\end{equation}
	where the contractions indicate the fermion propagators
	\begin{equation}\label{eq:ferm_prop}
	\bcontraction{}{\psi}{_i(t)\,}{\psi} \psi_i(t)\,\bpsi_j(t')\= \im\,S_{ij}(t,t') \ ,\quad\with
	\bigl(\im\delta_{ik}\sfrac{\diff}{\diff t}- V_{ik}\bigr)S_{kj}(t,t')\=\delta(t{-}t')\delta_{ij}\ ,
	\end{equation}
	and $\bcontraction{}{\bpsi}{_i(t)\,}{\psi} \bpsi_i(t)\,\psi_j(t')= -\im S_{ij}(t',t)$.
	Hence, we can write
	\begin{equation}
	R_g(g,\theta) \= \sfrac{1}{2}\int\!\diff t\ \diff t'\ (\partial_g V_i)(t)\ 
	\bigl[(1{+}\theta)S_{ij}(t',t)\;+\;(1{-}\theta)S_{ij}(t,t')\bigr]\ \sfrac{\delta}{\delta x_j(t')}\ .
	\end{equation}
	It is useful to define
	\begin{equation}
	\theta^{\pm}\ :=\ \sfrac{1}{2}(1\pm\theta)\ ,
	\end{equation}
	which allows us to write the $g$-flow operator compactly as
	\begin{equation}\label{eq:g_flow_op}
		\begin{aligned}
			R_g(g,\theta)\=\theta^+R_g^+(g)\;+\;\theta^-\,R_g^-(g) \quad\with R_g^+(g) &\= 
			\int\!\diff t\ \diff t'\ (\partial_g V_i)(t)\  S_{ij}(t',t)\ \sfrac{\delta}{\delta x_j(t')}\\ \quad\text{and}\quad R_g^-(g) &\= 
			\int\!\diff t\ \diff t'\ (\partial_g V_i)(t)\  S_{ij}(t,t')\ \sfrac{\delta}{\delta x_j(t')}\ .
		\end{aligned}
	\end{equation}
	
	The $\theta$-flow operator can be derived similarly. Employing \eqref{eq:theta_derivative} it becomes
	\begin{equation} \label{eq:theta_flow}
	\begin{aligned}
	R_\theta^\beta(g)\= 
	\sfrac{\im}{2}\,\bcontraction{}{\bar{\Delta}}{^\beta}{\delta} \bar{\Delta}^\beta \delta\,-\,
	\sfrac{\im}{2}\,\bcontraction{}{\Delta}{^\beta}{\bar{\delta}} \Delta^\beta \bar{\delta} \=
	\sfrac{\beta}{2} &\int\!\diff t\ \diff t'\ V_i(t)\,[S_{ij}(t',t)-S_{ij}(t,t')]\,\sfrac{\delta}{\delta x_j(t')}\\ \,+\ 
	\sfrac{\beta{-}1}{2} &\int\!\diff t\ \diff t'\ \im\dot{x}_i(t)\; [S_{ij}(t',t)+S_{ij}(t,t')]\,\sfrac{\delta}{\delta x_j(t')}\ ,
	\end{aligned}
	\end{equation}
	indicating the ambiguity~$\beta$. Note that it does not depend on~$\theta$.
	In Appendix \ref{app:n_proofs} we prove that $R_g$ \eqref{eq:g_flow_op} and $R_\theta^\beta$ \eqref{eq:theta_flow} 
	satisfy their respective free-action and determinant-matching condition~\eqref{eq:cf_cond12}. 

	\subsection{Free massive theory}\label{subsec:massive}
	Here, we consider the simple one-dimensional superpotential
	\begin{equation} \label{eq:harmonicV}
		V\=\sfrac12 mx^2\ ,
	\end{equation}
	giving rise to a free, yet massive Lagrangian
	\begin{equation}
		\Lag_0\=\sfrac{1}{2}\dot{x}^2-\sfrac12 m^2x^2+\bpsi\bigl[\im\sfrac{\diff}{\diff t}-m\bigr]\psi+\im\theta m\,x\dot{x}\ .
	\end{equation}
	In this case, the fermion propagators can be read off from
	\begin{equation}
		\bigl[\im\sfrac{\diff}{\diff t}- m\bigr]S_0(t,t')\=\delta(t{-}t') \qquad\Rightarrow\qquad
		\bcontraction{}{\psi}{(t)\,}{\psi} \psi(t)\,\bpsi(t')\=\im S_0(t,t')\=\tfrac{1}{2}\,\textrm{sgn}(t{-}t')\,\ep^{-\im m(t-t')}
	\end{equation}
	(with an integration constant fixed by antisymmetry for $m{=}0$), and the flow operators take the form
	\begin{equation} \label{eq:massive_flow}
		\begin{aligned}
		R_m(m,\theta)&\=\theta^+R_m^+(m)\,+\,\theta^-\,R_m^-(m)\ ,\quad\with 
		R_m^+(m) \= \int\!\diff t\ \diff t'\ x(t)\ S_0(t',t)\ \sfrac{\delta}{\delta x(t')}\\
		&\hspace{4.7cm}\quad\text{and}\quad R_m^-(m) \= \int\!\diff t\ \diff t'\ x(t)\ S_0(t,t')\ \sfrac{\delta}{\delta x(t')}\ ,\\
		R^{\beta}_{\theta}(m)&\=\sfrac{2\beta{-}1}{2}m \int\!\diff t\ \diff t'\ x(t)\ [S_0(t',t)-S_0(t,t')]\ \sfrac{\delta}{\delta x(t')}
		\=\sfrac{2\beta{-}1}{2}\,m\,[R_m^+(m)-R_m^-(m)]\ ,
		\end{aligned}
	\end{equation}
	where for the $\theta$~flow we used integration by parts making use of the fact 
	that $V\bigl(x(t)\bigr)\big|_{-\infty}^{+\infty}=0$ for finite-action trajectories in the harmonic potential~\eqref{eq:harmonicV}.
	Note that for $\beta{=}\frac12$ the $\theta$~flow can be made to vanish.

	We find it convenient to Fourier transform to frequency space, using the convention
	\begin{equation}
		x(t)\=\int\!\sfrac{\diff \omega}{2\pi}\ \ep^{\im\omega t}\ \x(\omega)\quad\Rightarrow\quad 
		\sfrac{\delta}{\delta x(t)}\=\int\!\diff\omega\ \ep^{-\im\omega t}\ \sfrac{\delta}{\delta \x(\omega)}\ .
	\end{equation}
	The fermion propagators get transformed as follows,
	\begin{equation}
		S_0(t,t')\=\int\!\sfrac{\diff \omega}{2\pi}\ \tS_0(\omega)\ \ep^{\im\omega (t-t')}
		\qquad\Rightarrow\qquad \tS_0(\omega)\=-\sfrac{1}{\omega+m}\ .
	\end{equation}
	The flow operators \eqref{eq:massive_flow} then are expressed via
	\begin{equation}
		R_m^\pm(m)\= \int\!\sfrac{\diff\omega}{2\pi}\ \x(\omega)\ \tS_0(\mp\omega)\ \sfrac{\delta}{\delta \x(\omega)}\ ,
	\end{equation}
	and their action on $\x(\omega)$ is simply multiplicative,
	\begin{equation}
		R_m(m,\theta)\  \x(\omega) \= \sfrac{m-\theta\,\omega}{\omega^2-m^2}\ \x(\omega) \quad\und\quad
		R^\beta_\theta(m)\  \x(\omega)\= (1{-}2\beta)\sfrac{m\;\omega}{\omega^2-m^2}\ \x(\omega)\ .
	\end{equation}

	\noindent\textbf{Nicolai maps.\ }
	Since in the free massive model, all flow operators are seen to commute with one another, the path ordering in the universal formula
	\eqref{eq:po_exp} (or its expanded form \eqref{eq:po_exp_exp}) trivializes.\footnote{
	    Still, the one-form $R(m,\theta)$ is only weakly flat since $\pa_\theta R_m-\pa_m R_\theta^\beta$ does not vanish.}
	Therefore, given an arbitrary path
	\begin{equation}
		h(s) \= \bigl(m(s),\theta(s)\bigr) \quad\with \bigl(m(0),\theta(0)\bigr)=(0,0) \und \bigl(m(1),\theta(1)\bigr)=(m,\theta)\ ,
	\end{equation}
	we can derive an explicit expression for the Nicolai map or its inverse
	\begin{equation} \label{eq:mass_inv_m_th}
	\begin{aligned}
		T^{-1}_{m,\theta}[h]\ \x(\omega) &\= \exp\Bigl\{\smallint_0^1\diff{s}\ 
		\bigl[ m'(s)\,R_m\bigl(m(s),\theta(s)\bigr)\ +\ \theta'(s)\,R^\beta_\theta\bigl(m(s),\theta(s)\bigr) \bigr]\Bigr\}\ \x(\omega) \\
		&\= \exp\Bigl\{\smallint_0^m \diff{\mu}\ R_m\bigl(\mu,\theta(\mu)\bigr)\Bigr\}\ 
		\exp\Bigl\{\smallint_0^\theta \diff{\vartheta}\ R_\theta^\beta\bigl(m(\vartheta),\vartheta\bigr)\Bigr\}\ \x(\omega) \\
		&\= \exp\Bigl\{\smallint_0^m \diff{\mu}\ \tfrac{\mu}{\omega^2-\mu^2} \Bigr\}\
		\exp\Bigl\{-\omega \smallint_0^m \diff{\mu}\ \tfrac{\theta(\mu)}{\omega^2-\mu^2} \Bigr\}\
		\exp\Bigl\{(1{-}2\beta)\,\omega \smallint_0^\theta \diff{\vartheta}\ \tfrac{m(\vartheta)}{\omega^2-m(\vartheta)^2}\Bigr\}\ \x(\omega) \\
		&\= \sqrt{\tfrac{\omega^2}{\omega^2-m^2}}\ \ep^{{}\;\omega\;\theta\,f(\omega^2)}\ \x(\omega)
	\end{aligned}
	\end{equation}
	for some function~$f$ depending on the contour,
	where after the first line we assumed, for simplicity, that the parametrizations $m(s)$ and $\theta(s)$ are monotonous, so that 
	we may pass to a global `coupling-coordinate parametrization' $\bigl(\mu,\theta(\mu)\bigr)$ with $\mu\in[0,m]$ and
	$\bigl(m(\vartheta),\vartheta\bigr)$ with $\vartheta\in[0,\theta]$, respectively.\footnote{
	    If this is not the case, the path may be composed of piecewise monotonous parts.}
	
	It is now straightforward to evaluate this expression for the contour of choice. Let us do this for three cases.
	The sequential flow `first $m$ then~$\theta$' produces
	 \begin{equation}\label{eq:mass_inv_theta}
                T^{-1}_{m,\theta}[\dcontour]\ \x \=
		\exp\Bigl\{(1{-}2\beta)\smallint_0^\theta \diff{\vartheta}\ \tfrac{m\,\omega}{\omega^2-m^2}\Bigr\}\
		\exp\Bigl\{\smallint_0^m \diff{\mu}\ \tfrac{\mu}{\omega^2-\mu^2} \Bigr\}\ \x\=
		\sqrt{\sfrac{\omega^2}{\omega^2-m^2}}\ \ep^{(1{-}2\beta)\frac{\theta\,m\,\omega}{\omega^2-m^2}}\ \x\ .
        \end{equation}
	On the other hand, for `first~$\theta$ then~$m$' the $\theta$ flow is trivial, and we obtain
	\begin{equation}\label{eq:mass_inv_m}
		T^{-1}_{m,\theta}[\ucontour]\ \x \=
		\exp\bigl\{\smallint_{0}^{m}\diff \mu\ \sfrac{\mu-\theta\,\omega}{\omega^2-\mu^2}\bigr\}\ \x\=
		(1+\sfrac{m}{\omega})^{-\sfrac{1+\theta}{2}}(1-\sfrac{m}{\omega})^{-\sfrac{1-\theta}{2}}\ \x\ \=
		\sqrt{\sfrac{\omega^2}{\omega^2-m^2}}\ \bigl(\sfrac{\omega-m}{\omega+m}\bigr)^{\theta/2}\ \x\ .
	\end{equation}
	These two versions may be contrasted with the direct-contour flow, via $m(\vartheta)=\frac{m}{\theta}\vartheta$ and $\theta(\mu)=\frac{\theta}{m}\mu$:
	\begin{equation}\label{eq:mass_inv_m_th_direct}
                T^{-1}_{m,\theta}[\scontour]\ \x \=
		\exp\Bigl\{\bigl[(1-\tfrac{\theta}{m}\omega)+(1{-}2\beta)\tfrac{\theta}{m}\omega\bigr]\smallint_0^m \diff{\mu}\ \tfrac{\mu}{\omega^2-\mu^2} \Bigr\}\ \x\=
		\sqrt{\sfrac{\omega^2}{\omega^2-m^2}}\ \bigl(\sfrac{\omega^2}{\omega^2-m^2}\bigr)^{-\beta\sfrac{\theta}{m}\omega}\ \x\ .
        \end{equation}
	The obvious lesson is that the (inverse) Nicolai map really depends on the chosen path for the coupling flow. 
	There occur simplifications for $\beta=\frac12$ or~$0$, and for $\theta=0$.
	One may check though that the condition~\eqref{eq:collapse_cond} for contour independence is not met, 
	and thus it was expected to encounter Nicolai maps highly nonlinear in the coupling and depending on the integration contour.
	However, the contour~$\ucontour$ is special, since it trivializes the $\theta$~flow (performed at $m{=}0$)
	and hence yields the partial Nicolai map for the $m$~flow alone with $\theta$ fixed as an external parameter: 
	$T_{m,\theta}[\ucontour]=T_m(\theta)$. 
	And indeed, at the two special values $\theta=\pm1$, a drastic simplification occurs:
	\begin{equation}
		T^{-1}_m({\pm}1)\ \x(\omega) \= \tfrac{\omega}{\omega\pm m}\ \x(\omega)
		\qquad\Rightarrow\qquad
		T_m({\pm}1)\ \x(\omega) \= \bigl(1\pm\sfrac{m}{\omega}\bigr)\ \x(\omega)\ .
	\end{equation}
	Fourier-transforming back to the time domain, this implies that
	\begin{equation} \label{eq:Tmpoly}
		T_m(\theta{=}{\pm}1)\ \im\dot{x}(t) \= \im\dot{x}(t)\ \mp\ m\,x(t)
	\end{equation}
	which corraborates \eqref{eq:polyTdot} and just relates the massive to the massless fermion propagator.

	\noindent\textbf{Correlators.\ }
	Using the Nicolai map formalism, we can deduce from the massless boson propagator
	\begin{equation}
		\bigl\langle \x(\omega)\ \x(\omega')\bigr\rangle_{0,0}\= 2\pi \delta(\omega+\omega')\big/\omega^2
	\end{equation}
	the massive boson propagator
	\begin{equation}
		\bigl\langle \x(\omega)\ \x(\omega')\bigr\rangle_{m,\theta}\= 
		\bigl\langle T^{-1}_{m,\theta}[h] \x(\omega)\ T^{-1}_{m,\theta}[h] \x(\omega')\bigr\rangle_{0,0} 
		\= 2\pi\delta(\omega+\omega') \big/ (\omega^2-m^2)\ ,
	\end{equation}
	for any choice of the contour~$h$. Clearly, the dependencies on $\theta$ or $\beta$ of the Nicolai map cancel out in the correlator,
	due to energy conservation $\omega'=-\omega$.  This generalizes to arbitrary $N$-point functions via Wick's theorem.

	We end this subsection with two remarks. 
	Firstly, even for $\theta=\pm1$ the Nicolai map depends on the integration contour~$h$ when $\theta$ is a variable coupling. 
	However, when $\theta$ is fixed at the outset (no $\theta$ flow) then the collapse of the power series for $T$ occurs for any choice of~$h$, 
	as we shall argue in the end of this section. 
	Secondly, if one Fourier transforms back to the time domain, one can explicitly check the flatness condition \eqref{eq:flatness_cond} of $R_m$ and $R_\theta$. 
	We find that only the expectation value of the curvature vanishes, but not the curvature itself. 
	This is expected since we have demonstrated above that the full Nicolai map is path dependent. 
	However, this dependence can only affect terms that have a vanishing expectation value.
	All possible Nicolai maps lead to the same correlation functions.
	
	\subsection{Interacting theory}\label{subsec:interacting}
	As compared to the previous subsection, we now add a cubic term to the superpotential,
	\begin{equation}
		V\=\sfrac12 mx^2\,+\,\sfrac13 gx^3\ ,
	\end{equation}
	so that the Lagrangian becomes
	\begin{equation}
		\Lag\=\sfrac12 \dot x^2 - \sfrac12 m^2x^2-mgx^3-\sfrac12 g^2x^4+
		\bpsi\bigl[\im\sfrac{\diff}{\diff t}-m-2gx\bigr]\psi+\im\theta(mx+gx^2)\,\dot{x}\ .
	\end{equation}
	The fermion propagators follow from
	\begin{equation}\label{eq:ferm_prop_mod}
		\bigl[\im\sfrac{\diff}{\diff t}- m-2gx(t)\bigr]\,S(t,t')\=\delta(t{-}t')\ ,
	\end{equation}
	while the flow operators read
	\begin{equation}
		\begin{aligned}
		&R_g(g,m,\theta)\=\theta^+\,R_g^+(g,m)\,+\,\theta^-\,R_g^-(g,m)\ ,\\
		&R_m\!(g,m,\theta)\=\theta^+\,R_m^+(g,m)\,+\,\theta^-\,R_m^-(g,m)\ ,
		\end{aligned}
	\end{equation}
	with $R_g^\pm(g,m)$ and $R_m^\pm(g,m)$ as in \eqref{eq:g_flow_op} but for $D{=}1$.
	In the following, we will always first move to a finite $\theta$~value (at $g{=}m{=}0$) before turning on the other couplings
	along some contour~$h$ in $(g,m)$~space. This (and taking $\beta{=}1$) trivializes the $\theta$~flow, so we drop the $\theta$~subscript on~$T$. 
	The choice of $\theta=\pm1$ implies that \eqref{eq:collapse_one_coupling} holds which renders the full Nicolai map linear in $g$ and~$m$ for any path~$h$,
	\begin{equation} \label{eq:Tgmpoly}
		T_{g,m}[h,\theta{=}{\pm}1]\ \im\dot{x}(t)\=\im\dot{x}(t)\;\mp\;m\,x(t)\;\mp\;g\,x(t)^2\ ,
	\end{equation}
	and generates the bosonic Lagrangian via $\sfrac12 (T_{m,g}\dot{x})^2$ (free-action condition), whereas the inverse map never truncates. The generalization to an arbitrary superpotential $V$ is completely straightforward, which justifies \eqref{eq:polyTdot}.

	However, we find it more useful to start already with the massive free theory and work with the one-variable $g$-flow map, 
	since in this way we avoid the infrared divergences of the massless propagators. 
	Hence, we investigate the (inverse) partial map $T_g(m,\theta)$ at fixed $m$ and~$\theta$, which has no contour ambiguity any more. 
	Again, the map is linear in the variable coupling~$g$ for $\theta=\pm1$:
	\begin{equation}\label{eq:polynomial_inverse_map_th_pm}
	\begin{aligned}
		&T_g(m,\theta{=}{\pm}1)\ \im\dot{x}(t)\=\im\dot{x}(t)\;\,\mp\;g\,x(t)^2
		\;-\; gm\int\!\diff t'\ \tfrac{1}{2\im}\,\textrm{sgn}(t{-}t')\;\ep^{\pm\im m(t-t')} \ x(t')^2 \\
		\Leftrightarrow\qquad
		&T_g(m,\theta{=}{\pm}1)\ \bigl[\im\sfrac{\diff}{\diff t} \pm m\bigr]\,x(t) \= \bigl[\im\sfrac{\diff}{\diff t} \mp m\bigr]\,x(t)\;\mp\;g\,x(t)^2\ .
	\end{aligned}
	\end{equation}
	Indeed, composing this with $T_m(\theta{=}{\pm}1)$ from \eqref{eq:Tmpoly} for a sequential contour is consistent with \eqref{eq:Tgmpoly},
	\begin{equation}
		T_{g,m}[\ucontour,\theta{=}{\pm}1]\ \im\dot{x}(t) \= 
		T_g(m,\theta{=}{\pm}1)\ T_m(g{=}0,\theta{=}{\pm}1)\ \im\dot{x}(t) \= 
		\im\dot{x}(t)\;\mp\;m\,x(t)\;\mp\;g\,x(t)^2 \ .
	\end{equation}
	Here, it is also straightforward to verify the free-action condition.
	
	For the perturbative expansion of the inverse Nicolai map, we transform the $g$-flow operator to frequency space
	and expand \`a la~\eqref{eq:Rg_exp} to obtain the $O(g^{k-1})$ term
	\begin{equation}\label{eq:Rk_orders}
	\begin{aligned}
	\rR_k\=
	&\theta^+2^{k-1}\int\!\sfrac{\diff \nu_1}{2\pi}\cdots\sfrac{\diff \nu_{k+1}}{2\pi}\ 
	\x(\nu_1)[\x(-\nu_1+\nu_2)\tS_0(+\nu_2)]\cdots[\x(-\nu_k+\nu_{k+1})\tS_0(+\nu_{k+1})]\ \sfrac{\delta}{\delta \x(\nu_{k+1})}\\
	\ +\ 
	&\theta^-2^{k-1}\int\!\sfrac{\diff \nu_1}{2\pi}\cdots\sfrac{\diff \nu_{k+1}}{2\pi}\ 
	\x(\nu_1)[\x(-\nu_1+\nu_2)\tS_0(-\nu_2)]\cdots[\x(-\nu_k+\nu_{k+1})\tS_0(-\nu_{k+1})]\ \sfrac{\delta}{\delta \x(\nu_{k+1})}\ .
	\end{aligned}
	\end{equation}
	We now have all the ingredients for the map \cite{LR1}
	\begin{equation}
	\begin{aligned}
	T_g\x(\omega) &\= \x(\omega) \ -\ g\,\rR_1 \x(\omega) \ -\ \sfrac12g^2\bigl(\rR_2-\rR_1^2\bigr)\x(\omega)\ -\ 
	\sfrac16g^3\bigl(2\rR_3-2\rR_2\rR_1-\rR_1\rR_2+\rR_1^3\bigr)\x(\omega) \\[4pt]
	&\ -\ \sfrac{1}{24}g^4\bigl(6\rR_4-6\rR_3\rR_1-3\rR_2\rR_2+3\rR_2\rR_1^2-2\rR_1\rR_3+2\rR_1\rR_2\rR_1+\rR_1^2\rR_2-\rR_1^4\bigr)\x(\omega) \ +\ {\cal O}(g^5)
	\end{aligned}
	\end{equation}
	and its inverse
	\begin{equation}
	\begin{aligned}
	T_g^{-1}\x(\omega) &\= \x(\omega) \ +\ g\,\rR_1 \x(\omega) \ +\ \sfrac12g^2\bigl(\rR_2+\rR_1^2\bigr)\x(\omega)\ +\ 
	\sfrac16g^3\bigl(2\rR_3+2\rR_1\rR_2+\rR_2\rR_1+\rR_1^3\bigr)\x(\omega) \\[4pt]
	\ +&\ \sfrac{1}{24}g^4\bigl(6\rR_4+6\rR_1\rR_3+3\rR_2\rR_2+3\rR_1^2\rR_2+2\rR_3\rR_1+2\rR_1\rR_2\rR_1+\rR_2\rR_1^2+\rR_1^4\bigr)\x(\omega) \ +\ {\cal O}(g^5)\ .
	\end{aligned}
	\end{equation}
	Next, we will give an appropriate diagrammatic notation for the interacting theory. 
	In the following section, this will be employed to compactly write expressions for $T_g\x(\omega)$ and for $T_g^{-1}\x(\omega)$ to order $g^3$ 
	for arbitrary $\theta$ and to order $g^4$ for $\theta=\pm1$.
	
	\subsection{Nicolai rules}\label{subsec:Nicolai_rules}
 	We introduce a graphical Feynman-like notation, which we call Nicolai rules. 
	We find it useful to work in frequency space, dropping the tildes from here onward. This gives the following rules:
	\begin{itemize}
		\item External boson lines with a frequency $\omega$ give a factor $x(\omega)$.
		\item 	Free fermion propagators are
		\begin{equation}
		S_0(\omega)\quad\=\quad
		\begin{tikzpicture}[baseline={([yshift=-.1cm]current bounding box.center)}]
			\begin{feynman}
				\vertex (a);
				\vertex [right=of a] (b);
				\diagram* {
					(a) -- [fermion] (b),
				};
			\end{feynman}\;\end{tikzpicture}\quad \=\quad\frac{-1}{\omega+m}\ ,\qquad
		S_0(-\omega)\quad\=\quad\begin{tikzpicture}[baseline={([yshift=-.1cm]current bounding box.center)}]
			\begin{feynman}
				\vertex (a);
				\vertex [right=of a] (b);
				\diagram* {
					(a) -- [anti fermion] (b),
				};
			\end{feynman}\;\end{tikzpicture}\quad \=\quad\frac{1}{\omega-m}\ .
		\end{equation}
		\item Free boson propagators are
		\begin{equation}
		G_0(\omega)\quad=\quad\begin{tikzpicture}[baseline={([yshift=-.1cm]current bounding box.center)}]
			\begin{feynman}
				\vertex (a);
				\vertex [right=of a] (b);
				\diagram* {
					(a) -- [photon] (b),
				};
			\end{feynman}\;\end{tikzpicture}\quad \=\quad\frac{1}{\omega^2-m^2}\ .
		\end{equation}
		\item Vertices (implicit of order $g$) give factors
		\begin{equation}
		\begin{tikzpicture}[baseline={([yshift=-.1cm]current bounding box.center)}, scale=0.5, transform shape]
				\begin{feynman}
					\vertex (a);
					\vertex [right=of a] (b);
					\vertex [above left=of a] (c1);
					\vertex [below left=of a] (c2);
					\diagram* {
						(c1) -- [photon] (a),
						(c2) -- [photon] (a),
						(a) -- (b),
					};
				\end{feynman}\;\end{tikzpicture}\quad=\quad 1\ ,\qquad 
		\begin{tikzpicture}[baseline={([yshift=-.45cm]current bounding box.center)}, scale=0.5, transform shape]
				\begin{feynman}
					\vertex (a);
					\vertex [right=of a] (b);
					\vertex [right=of b] (c);
					\vertex [above=of b] (t);
					\diagram* {
						(a) -- (b) -- (c),
						(b) -- [photon] (t),
					};
				\end{feynman}\;\end{tikzpicture}\quad\=\quad 2\ ,\qquad 
		\begin{tikzpicture}[baseline={([yshift=-.45cm]current bounding box.center)}, scale=0.5, transform shape]
				\begin{feynman}
					\vertex (a);
					\vertex [right=of a] (b);
					\vertex [right=of b] (c);
					\vertex [above=of b] (t);
					\diagram* {
						(a) -- (b) -- (c),
						(b) -- (t),
					};
				\end{feynman}\;\end{tikzpicture}\quad=\quad 2\ .
		\end{equation}
		\item At every vertex, energy conservation is enforced. We take all frequencies to be oriented towards the root of the tree, which carries the frequency $\omega$ of the transformed field $T_g\, x(\omega)$. Each remaining frequency $\nu$ comes with an integral $\int \sfrac{\diff \nu}{2\pi}$.
		
	\end{itemize}
	Using these rules, we can represent the $g$-flow operator in a graphical notation (c.f.~\eqref{eq:Rk_orders})
	\begin{equation}
		\begin{aligned}
			R_g(g,m,\theta)\=&\Bigl\{
			\theta^+\begin{tikzpicture}[baseline={([yshift=-.1cm]current bounding box.center)}, scale=0.4, transform shape]
				\begin{feynman}
					\vertex (a);
					\vertex [right=of a] (c);
					\vertex [above left=of a] (d);
					\vertex [below left=of a] (e);
					\diagram* {
						(a) -> [fermion] (c),
						(a) -- [photon] (d),
						(a) -- [photon] (e),
					};
				\end{feynman}\;\end{tikzpicture}
			\ +\ \theta^-
			\begin{tikzpicture}[baseline={([yshift=-.1cm]current bounding box.center)}, scale=0.4, transform shape]
				\begin{feynman}
					\vertex (a);
					\vertex [right=of a] (c);
					\vertex [above left=of a] (d);
					\vertex [below left=of a] (e);
					\diagram* {
						(a) -> [anti fermion] (c),
						(a) -- [photon] (d),
						(a) -- [photon] (e),
					};
				\end{feynman}\;\end{tikzpicture}\Bigr\}
			\ +\ g\ \Bigl\{\theta^+\begin{tikzpicture}[baseline={([yshift=-.16cm]current bounding box.center)}, scale=0.4, transform shape]
				\begin{feynman}
					\vertex (a);
					\vertex [right=of a] (b);
					\vertex [right=of b] (c);
					\vertex [above left=of a] (d);
					\vertex [below left=of a] (e);
					\vertex [above=of b] (f);
					\diagram* {
						(a) -- [fermion] (b) -> [fermion] (c),
						(a) -- [photon] (d),
						(b) -- [photon] (f),
						(a) -- [photon] (e),
					};
				\end{feynman}\;\end{tikzpicture}
			\ +\ \theta^-
			\begin{tikzpicture}[baseline={([yshift=-.16cm]current bounding box.center)}, scale=0.4, transform shape]
				\begin{feynman}
					\vertex (a);
					\vertex [right=of a] (b);
					\vertex [right=of b] (c);
					\vertex [above left=of a] (d);
					\vertex [below left=of a] (e);
					\vertex [above=of b] (f);
					\diagram* {
						(a) -- [anti fermion] (b) -> [anti fermion] (c),
						(a) -- [photon] (d),
						(b) -- [photon] (f),
						(a) -- [photon] (e),
					};
				\end{feynman}\;\end{tikzpicture}\Bigr\}\\
			+\ &g^2\ \Bigl\{\theta^+
			\begin{tikzpicture}[baseline={([yshift=-.16cm]current bounding box.center)}, scale=0.4, transform shape]
				\begin{feynman}
					\vertex (a);
					\vertex [right=of a] (b);
					\vertex [right=of b] (c);
					\vertex [right=of c] (d);
					\vertex [above left=of a] (o1);
					\vertex [below left=of a] (o2);
					\vertex [above=of b] (be);
					\vertex [above=of c] (ce);
					\diagram* {
						(a) -- [fermion] (b) -- [fermion] (c) -> [fermion] (d),
						(b) -- [photon] (be),
						(c) -- [photon] (ce),
						(a) -- [photon] (o1),
						(a) -- [photon] (o2),
					};
				\end{feynman}\;
			\end{tikzpicture}\ +\ \theta^-
			\begin{tikzpicture}[baseline={([yshift=-.16cm]current bounding box.center)}, scale=0.4, transform shape]
				\begin{feynman}
					\vertex (a);
					\vertex [right=of a] (b);
					\vertex [right=of b] (c);
					\vertex [right=of c] (d);
					\vertex [above left=of a] (o1);
					\vertex [below left=of a] (o2);
					\vertex [above=of b] (be);
					\vertex [above=of c] (ce);
					\diagram* {
						(a) -- [anti fermion] (b) -- [anti fermion] (c) -> [anti fermion] (d),
						(b) -- [photon] (be),
						(c) -- [photon] (ce),
						(a) -- [photon] (o1),
						(a) -- [photon] (o2),
					};
				\end{feynman}\;
			\end{tikzpicture}\Bigr\}\ +\ \mathcal{O}(g^4)\ ,
		\end{aligned}
	\end{equation}
	where the arrows at the end of the fermion lines indicate a functional derivative with respect to $x$. 
	For arbitrary $\theta$, the map contains all possible combinations of fermion propagators in each topology:
	\begin{equation}\label{eq:map_interacting_theta}
		\begin{aligned}
			T_g(m,\theta)\ x(\omega)\=& x(\omega)
			\ -\ g\ \Bigl\{
			\theta^+\begin{tikzpicture}[baseline={([yshift=-.1cm]current bounding box.center)}, scale=0.4, transform shape]
					\begin{feynman}
						\vertex (a);
						\vertex [right=of a] (c);
						\vertex [above left=of a] (d);
						\vertex [below left=of a] (e);
						\diagram* {
							(a) -- [fermion] (c),
							(a) -- [photon] (d),
							(a) -- [photon] (e),
						};
					\end{feynman}\;\end{tikzpicture}
			\ +\ \theta^-
			\begin{tikzpicture}[baseline={([yshift=-.1cm]current bounding box.center)}, scale=0.4, transform shape]
					\begin{feynman}
						\vertex (a);
						\vertex [right=of a] (c);
						\vertex [above left=of a] (d);
						\vertex [below left=of a] (e);
						\diagram* {
							(c) -- [fermion] (a),
							(a) -- [photon] (d),
							(a) -- [photon] (e),
						};
					\end{feynman}\;\end{tikzpicture}
			\Bigr\}\\ \ -\ & \sfrac{g^2}{2}\ \theta^+\theta^-\Bigl\{\begin{tikzpicture}[baseline={([yshift=-.16cm]current bounding box.center)}, scale=0.4, transform shape]
					\begin{feynman}
						\vertex (a);
						\vertex [right=of a] (b);
						\vertex [right=of b] (c);
						\vertex [above left=of a] (d);
						\vertex [below left=of a] (e);
						\vertex [above=of b] (f);
						\diagram* {
							(a) -- [fermion] (b) -- [fermion] (c),
							(b) -- [photon] (f),
							(a) -- [photon] (d),
							(a) -- [photon] (e),
						};
					\end{feynman}\;
			\end{tikzpicture}\ +\ \begin{tikzpicture}[baseline={([yshift=-.16cm]current bounding box.center)}, scale=0.4, transform shape]
					\begin{feynman}
						\vertex (a);
						\vertex [right=of a] (b);
						\vertex [right=of b] (c);
						\vertex [above left=of a] (d);
						\vertex [below left=of a] (e);
						\vertex [above=of b] (f);
						\diagram* {
							(c) -- [fermion] (b) -- [fermion] (a),
							(b) -- [photon] (f),
							(a) -- [photon] (d),
							(a) -- [photon] (e),
						};
					\end{feynman}\;
			\end{tikzpicture}\ -\ \begin{tikzpicture}[baseline={([yshift=-.16cm]current bounding box.center)}, scale=0.4, transform shape]
					\begin{feynman}
						\vertex (a);
						\vertex [right=of a] (b);
						\vertex [right=of b] (c);
						\vertex [above left=of a] (d);
						\vertex [below left=of a] (e);
						\vertex [above=of b] (f);
						\diagram* {
							(a) -- [fermion] (b),
							(c) -- [fermion] (b),
							(b) -- [photon] (f),
							(a) -- [photon] (d),
							(a) -- [photon] (e),
						};
					\end{feynman}\;
			\end{tikzpicture}\ -\ \begin{tikzpicture}[baseline={([yshift=-.16cm]current bounding box.center)}, scale=0.4, transform shape]
					\begin{feynman}
						\vertex (a);
						\vertex [right=of a] (b);
						\vertex [right=of b] (c);
						\vertex [above left=of a] (d);
						\vertex [below left=of a] (e);
						\vertex [above=of b] (f);
						\diagram* {
							(b) -- [fermion] (a),
							(b) -- [fermion] (c),
							(b) -- [photon] (f),
							(a) -- [photon] (d),
							(a) -- [photon] (e),
						};
					\end{feynman}\;
			\end{tikzpicture}\Bigr\}\\ \ -\ &\sfrac{g^3}{6}\theta^+\theta^-\Bigl\{(1{+}\theta^-)\begin{tikzpicture}[baseline={([yshift=-.16cm]current bounding box.center)}, scale=0.4, transform shape]
					\begin{feynman}
						\vertex (a);
						\vertex [right=of a] (b);
						\vertex [right=of b] (c);
						\vertex [right=of c] (d);
						\vertex [above left=of a] (o1);
						\vertex [below left=of a] (o2);
						\vertex [above=of b] (be);
						\vertex [above=of c] (ce);
						\diagram* {
							(a) -- [fermion] (b) -- [fermion] (c) -- [fermion] (d),
							(b) -- [photon] (be),
							(c) -- [photon] (ce),
							(a) -- [photon] (o1),
							(a) -- [photon] (o2),
						};
					\end{feynman}\;
			\end{tikzpicture}
		\ -\ \theta^+
		\begin{tikzpicture}[baseline={([yshift=-.4cm]current bounding box.center)}, scale=0.4, transform shape]
		\begin{feynman}
		\vertex (a);
		\vertex [right=of a] (b);
		\vertex [right=of b] (c);
		\vertex [above left=of a] (d);
		\vertex [below left=of a] (e);
		\vertex [above=of b] (f);
		\vertex [above right=of f] (f1);
		\vertex [above left=of f] (f2);
		\diagram* {
			(a) -- [fermion] (b) -- [fermion] (c),
			(b) -- [anti fermion] (f),
			(f) -- [photon] (f1),
			(f) -- [photon] (f2),
			(a) -- [photon] (d),
			(a) -- [photon] (e),
		};
		\end{feynman}\;
		\end{tikzpicture}\\
		&\ \ -\ 	\theta^-\begin{tikzpicture}[baseline={([yshift=-.16cm]current bounding box.center)}, scale=0.4, transform shape]
					\begin{feynman}
						\vertex (a);
						\vertex [right=of a] (b);
						\vertex [right=of b] (c);
						\vertex [right=of c] (d);
						\vertex [above left=of a] (o1);
						\vertex [below left=of a] (o2);
						\vertex [above=of b] (be);
						\vertex [above=of c] (ce);
						\diagram* {
							(a) -- [anti fermion] (b) -- [fermion] (c) -- [fermion] (d),
							(b) -- [photon] (be),
							(c) -- [photon] (ce),
							(a) -- [photon] (o1),
							(a) -- [photon] (o2),
						};
					\end{feynman}\;
			\end{tikzpicture}
			\ -\ (1{+}\theta^+)
			\begin{tikzpicture}[baseline={([yshift=-.16cm]current bounding box.center)}, scale=0.4, transform shape]
					\begin{feynman}
						\vertex (a);
						\vertex [right=of a] (b);
						\vertex [right=of b] (c);
						\vertex [right=of c] (d);
						\vertex [above left=of a] (o1);
						\vertex [below left=of a] (o2);
						\vertex [above=of b] (be);
						\vertex [above=of c] (ce);
						\diagram* {
							(a) -- [anti fermion] (b) -- [anti fermion] (c) -- [fermion] (d),
							(b) -- [photon] (be),
							(c) -- [photon] (ce),
							(a) -- [photon] (o1),
							(a) -- [photon] (o2),
						};
					\end{feynman}\;
			\end{tikzpicture}
			\ +\ \theta^+
			\begin{tikzpicture}[baseline={([yshift=-.16cm]current bounding box.center)}, scale=0.4, transform shape]
					\begin{feynman}
						\vertex (a);
						\vertex [right=of a] (b);
						\vertex [right=of b] (c);
						\vertex [right=of c] (d);
						\vertex [above left=of a] (o1);
						\vertex [below left=of a] (o2);
						\vertex [above=of b] (be);
						\vertex [above=of c] (ce);
						\diagram* {
							(a) -- [fermion] (b) -- [anti fermion] (c) -- [fermion] (d),
							(b) -- [photon] (be),
							(c) -- [photon] (ce),
							(a) -- [photon] (o1),
							(a) -- [photon] (o2),
						};
					\end{feynman}\;
			\end{tikzpicture}\\
			&\ \ +\ (\theta^+{-}\theta^-)
			\begin{tikzpicture}[baseline={([yshift=-.4cm]current bounding box.center)}, scale=0.4, transform shape]
					\begin{feynman}
						\vertex (a);
						\vertex [right=of a] (b);
						\vertex [right=of b] (c);
						\vertex [above left=of a] (d);
						\vertex [below left=of a] (e);
						\vertex [above=of b] (f);
						\vertex [above right=of f] (f1);
						\vertex [above left=of f] (f2);
						\diagram* {
							(a) -- [fermion] (b) -- [fermion] (c),
							(b) -- [fermion] (f),
							(f) -- [photon] (f1),
							(f) -- [photon] (f2),
							(a) -- [photon] (d),
							(a) -- [photon] (e),
						};
					\end{feynman}\;
			\end{tikzpicture}
			\ +\ \theta^- \begin{tikzpicture}[baseline={([yshift=-.4cm]current bounding box.center)}, scale=0.4, transform shape]
					\begin{feynman}
						\vertex (a);
						\vertex [right=of a] (b);
						\vertex [right=of b] (c);
						\vertex [above left=of a] (d);
						\vertex [below left=of a] (e);
						\vertex [above=of b] (f);
						\vertex [above right=of f] (f1);
						\vertex [above left=of f] (f2);
						\diagram* {
							(a) -- [anti fermion] (b) -- [fermion] (c),
							(b) -- [fermion] (f),
							(f) -- [photon] (f1),
							(f) -- [photon] (f2),
							(a) -- [photon] (d),
							(a) -- [photon] (e),
						};
					\end{feynman}\;
			\end{tikzpicture}
			\ +\ \bigl(\theta^+\leftrightarrow\theta^-,\Sr\leftrightarrow\Sl\bigr)\Bigr\}\ +\ \mathcal{O}(g^4)\ ,
		\end{aligned}
	\end{equation}
	where $\Sr\leftrightarrow\Sl$ indicates a reversal of the arrow on each fermion propagator.
	From this representation, it is easy to see that for $\theta=\pm 1$ the map truncates after ${\cal O}(g)$, 
	since all higher-order terms are multiplied with $\theta^+\theta^-= \theta^+(1{-}\theta^+) = \theta^-(1{-}\theta^-)$. 
	In these cases, $T_g$ becomes linear in $g$, polynomial (quadratic) in $x$, and only one of the elementary propagators is present. 
	The latter fact implies the identity $\rR_k \rR_1 x=\rR_{k+1}x$~\eqref{eq:uniqueness_cond} so that we find
	\begin{equation}\label{eq:polynomial_Tg}
		T_g(m,{\pm}1)\;x\ \=\ x\ -\ g\,\rR_1\;x\ \=\ x\ -\ g\begin{tikzpicture}[baseline={([yshift=-.1cm]current bounding box.center)}, scale=0.4,transform shape]
				\begin{feynman}
					\vertex (a);
					\vertex [right=of a] (c);
					\vertex [above left=of a] (d);
					\vertex [below left=of a] (e);
					\vertex [above=of b] (f);
					\diagram* {
						(a) -- [] (c),
						(a) -- [photon] (d),
						(a) -- [photon] (e),
					};
				\end{feynman}\;
		\end{tikzpicture}\ ,
	\end{equation}
	with an implicit arrow on the fermion propagator to the right (or left) for $\theta=+1$ (or $-1$). 
	We argue that, even for an arbitrary superpotential $V$ which depends on multiple couplings $g$, 
	the identity~\eqref{eq:collapse_cond} must hold whenever $\theta=\pm 1$, so that the power series truncates for any integration contour in~\eqref{eq:po_exp} 
	(see the end of Section~\ref{sec:nmap}). If one inverts \eqref{eq:polynomial_Tg} iteratively, one finds the expansion
	\begin{equation}\label{eq:inv_map_interacting_pm}
		\begin{aligned}
			T_g^{-1}(m,{\pm}1)\;x(\omega)\=& x(\omega)
			\ +\ g\begin{tikzpicture}[baseline={([yshift=-.1cm]current bounding box.center)}, scale=0.4, transform shape]
					\begin{feynman}
						\vertex (a);
						\vertex [right=of a] (c);
						\vertex [above left=of a] (d);
						\vertex [below left=of a] (e);
						\vertex [above=of b] (f);
						\diagram* {
							(a) -- [] (c),
							(a) -- [photon] (d),
							(a) -- [photon] (e),
						};
					\end{feynman}\;
			\end{tikzpicture}\ +\  g^2
			\begin{tikzpicture}[baseline={([yshift=-.16cm]current bounding box.center)}, scale=0.4, transform shape]
					\begin{feynman}
						\vertex (a);
						\vertex [right=of a] (b);
						\vertex [right=of b] (c);
						\vertex [above left=of a] (d);
						\vertex [below left=of a] (e);
						\vertex [above=of b] (f);
						\diagram* {
							(a) -- (b) -- [] (c),
							(b) -- [photon] (f),
							(a) -- [photon] (d),
							(a) -- [photon] (e),
						};
					\end{feynman}\;
			\end{tikzpicture}\ +\ g^3\Bigl\{
			\begin{tikzpicture}[baseline={([yshift=-.16cm]current bounding box.center)}, scale=0.4, transform shape]
					\begin{feynman}
						\vertex (a);
						\vertex [right=of a] (b);
						\vertex [right=of b] (c);
						\vertex [right=of c] (z);
						\vertex [above left=of a] (d);
						\vertex [below left=of a] (e);
						\vertex [above=of b] (f);
						\vertex [above=of c] (y);
						\diagram* {
							(a) --  (b) -- (c) -- [] (z),
							(b) -- [photon] (f),
							(a) -- [photon] (d),
							(a) -- [photon] (e),
							(c) -- [photon] (y),
						};
					\end{feynman}\;
			\end{tikzpicture}\ +\ \sfrac12
			\begin{tikzpicture}[baseline={([yshift=-.4cm]current bounding box.center)}, scale=0.4, transform shape]
					\begin{feynman}
						\vertex (a);
						\vertex [right=of a] (b);
						\vertex [right=of b] (c);
						\vertex [above left=of a] (d);
						\vertex [below left=of a] (e);
						\vertex [above=of b] (f);
						\vertex [above right=of f] (f1);
						\vertex [above left=of f] (f2);
						\diagram* {
							(a) -- (b) -- [] (c),
							(b) -- (f),
							(f) -- [photon] (f1),
							(f) -- [photon] (f2),
							(a) -- [photon] (d),
							(a) -- [photon] (e),
						};
					\end{feynman}\;
			\end{tikzpicture}\Bigr\}\\
			\ &+\ g^4\Bigl\{\begin{tikzpicture}[baseline={([yshift=-.16cm]current bounding box.center)}, scale=0.4, transform shape]
					\begin{feynman}
						\vertex (a);
						\vertex [right=of a] (b);
						\vertex [right=of b] (c);
						\vertex [right=of c] (d);
						\vertex [right=of d] (e);
						\vertex [above left=of a] (a1);
						\vertex [below left=of a] (a2);
						\vertex [above=of b] (be);
						\vertex [above=of c] (ce);
						\vertex [above=of d] (de);
						\diagram* {
							(a) -- (b) -- (c) -- (d) -- (e),
							(b) -- [photon] (be),
							(c) -- [photon] (ce),
							(d) -- [photon] (de),
							(a) -- [photon] (a1),
							(a) -- [photon] (a2),
						};
					\end{feynman}\;
			\end{tikzpicture}\ +\ 
			\begin{tikzpicture}[baseline={([yshift=-.45cm]current bounding box.center)}, scale=0.4, transform shape]
					\begin{feynman}
						\vertex (a);
						\vertex [right=of a] (b);
						\vertex [right=of b] (c);
						\vertex [right=of c] (d);
						\vertex [above left=of a] (a1);
						\vertex [below left=of a] (a2);
						\vertex [above=of c] (c1);
						\vertex [above right=of c1] (c2);
						\vertex [above left=of c1] (c3);
						\vertex [above=of b] (be);
						\diagram* {
							(a) -- (b) -- (c) -- (d),
							(c) -- (c1) -- [photon] (c2),
							(c1) -- [photon] (c3),
							(b) -- [photon] (be),
							(a) -- [photon] (a1),
							(a) -- [photon] (a2),
						};
					\end{feynman}\;
			\end{tikzpicture}\ +\ \sfrac12
			\begin{tikzpicture}[baseline={([yshift=-.45cm]current bounding box.center)}, scale=0.4, transform shape]
					\begin{feynman}
						\vertex (a);
						\vertex [right=of a] (b);
						\vertex [right=of b] (c);
						\vertex [right=of c] (d);
						\vertex [above left=of a] (a1);
						\vertex [below left=of a] (a2);
						\vertex [above=of b] (b1);
						\vertex [above right=of b1] (b2);
						\vertex [above left=of b1] (b3);
						\vertex [above=of c] (ce);
						\diagram* {
							(a) -- (b) -- (c) -- (d),
							(b) -- (b1) -- [photon] (b2),
							(b1) -- [photon] (b3),
							(c) -- [photon] (ce),
							(a) -- [photon] (a1),
							(a) -- [photon] (a2),
						};
					\end{feynman}\;
			\end{tikzpicture}\Bigr\}\ +\ \mathcal{O}(g^5)\ .
		\end{aligned}
	\end{equation}
	Interestingly, a recursive argument shows that every topology comes with a weight of unity times a symmetry factor, 
	see e.g.~the diagrams with the coefficient $\tfrac12$ above.
	For arbitrary $\theta$, the first few terms of the inverse map are
	\begin{equation}\label{eq:inv_map_interacting_theta}
		\begin{aligned}
			T_g^{-1}&(m,\theta)\;x(\omega)\= x(\omega)
			\ +\ g\ \Bigl\{
			\theta^+\begin{tikzpicture}[baseline={([yshift=-.1cm]current bounding box.center)}, scale=0.4, transform shape]
					\begin{feynman}
						\vertex (a);
						\vertex [right=of a] (c);
						\vertex [above left=of a] (d);
						\vertex [below left=of a] (e);
						\diagram* {
							(a) -- [fermion] (c),
							(a) -- [photon] (d),
							(a) -- [photon] (e),
						};
					\end{feynman}\;\end{tikzpicture}
			\ +\ \theta^-
			\begin{tikzpicture}[baseline={([yshift=-.1cm]current bounding box.center)}, scale=0.4, transform shape]
					\begin{feynman}
						\vertex (a);
						\vertex [right=of a] (c);
						\vertex [above left=of a] (d);
						\vertex [below left=of a] (e);
						\diagram* {
							(c) -- [fermion] (a),
							(a) -- [photon] (d),
							(a) -- [photon] (e),
						};
					\end{feynman}\;\end{tikzpicture}
			\Bigr\}\\ \ +\ & \sfrac{g^2}{2}\Bigl\{\theta^+(1{+}\theta^+)\begin{tikzpicture}[baseline={([yshift=-.16cm]current bounding box.center)}, scale=0.4, transform shape]
					\begin{feynman}
						\vertex (a);
						\vertex [right=of a] (b);
						\vertex [right=of b] (c);
						\vertex [above left=of a] (d);
						\vertex [below left=of a] (e);
						\vertex [above=of b] (f);
						\diagram* {
							(a) -- [fermion] (b) -- [fermion] (c),
							(b) -- [photon] (f),
							(a) -- [photon] (d),
							(a) -- [photon] (e),
						};
					\end{feynman}\;
			\end{tikzpicture}\ +\ \theta^-(1{+}\theta^-)\begin{tikzpicture}[baseline={([yshift=-.16cm]current bounding box.center)}, scale=0.4, transform shape]
					\begin{feynman}
						\vertex (a);
						\vertex [right=of a] (b);
						\vertex [right=of b] (c);
						\vertex [above left=of a] (d);
						\vertex [below left=of a] (e);
						\vertex [above=of b] (f);
						\diagram* {
							(c) -- [fermion] (b) -- [fermion] (a),
							(b) -- [photon] (f),
							(a) -- [photon] (d),
							(a) -- [photon] (e),
						};
					\end{feynman}\;
			\end{tikzpicture}\ +\ \theta^+\theta^-\Bigl(\begin{tikzpicture}[baseline={([yshift=-.16cm]current bounding box.center)}, scale=0.4, transform shape]
					\begin{feynman}
						\vertex (a);
						\vertex [right=of a] (b);
						\vertex [right=of b] (c);
						\vertex [above left=of a] (d);
						\vertex [below left=of a] (e);
						\vertex [above=of b] (f);
						\diagram* {
							(a) -- [fermion] (b),
							(c) -- [fermion] (b),
							(b) -- [photon] (f),
							(a) -- [photon] (d),
							(a) -- [photon] (e),
						};
					\end{feynman}\;
			\end{tikzpicture}\ +\ \begin{tikzpicture}[baseline={([yshift=-.16cm]current bounding box.center)}, scale=0.4, transform shape]
					\begin{feynman}
						\vertex (a);
						\vertex [right=of a] (b);
						\vertex [right=of b] (c);
						\vertex [above left=of a] (d);
						\vertex [below left=of a] (e);
						\vertex [above=of b] (f);
						\diagram* {
							(b) -- [fermion] (a),
							(b) -- [fermion] (c),
							(b) -- [photon] (f),
							(a) -- [photon] (d),
							(a) -- [photon] (e),
						};
					\end{feynman}\;
			\end{tikzpicture}\Bigr)\Bigr\}\\ \ +\ &\sfrac{g^3}{6}\Bigl\{\theta^+(1{+}\theta^+)(2{+}\theta^+)\begin{tikzpicture}[baseline={([yshift=-.16cm]current bounding box.center)}, scale=0.4, transform shape]
					\begin{feynman}
						\vertex (a);
						\vertex [right=of a] (b);
						\vertex [right=of b] (c);
						\vertex [right=of c] (d);
						\vertex [above left=of a] (o1);
						\vertex [below left=of a] (o2);
						\vertex [above=of b] (be);
						\vertex [above=of c] (ce);
						\diagram* {
							(a) -- [fermion] (b) -- [fermion] (c) -- [fermion] (d),
							(b) -- [photon] (be),
							(c) -- [photon] (ce),
							(a) -- [photon] (o1),
							(a) -- [photon] (o2),
						};
					\end{feynman}\;
			\end{tikzpicture}
			\ +\ \theta^+\theta^+(2{+}\theta^+)
			\begin{tikzpicture}[baseline={([yshift=-.4cm]current bounding box.center)}, scale=0.4, transform shape]
				\begin{feynman}
					\vertex (a);
					\vertex [right=of a] (b);
					\vertex [right=of b] (c);
					\vertex [above left=of a] (d);
					\vertex [below left=of a] (e);
					\vertex [above=of b] (f);
					\vertex [above right=of f] (f1);
					\vertex [above left=of f] (f2);
					\diagram* {
						(a) -- [fermion] (b) -- [fermion] (c),
						(b) -- [anti fermion] (f),
						(f) -- [photon] (f1),
						(f) -- [photon] (f2),
						(a) -- [photon] (d),
						(a) -- [photon] (e),
					};
				\end{feynman}\;
			\end{tikzpicture}\\
			&\ \ +\ \theta^+\theta^-\Bigl[(2{+}\theta^+)\begin{tikzpicture}[baseline={([yshift=-.16cm]current bounding box.center)}, scale=0.4, transform shape]
					\begin{feynman}
						\vertex (a);
						\vertex [right=of a] (b);
						\vertex [right=of b] (c);
						\vertex [right=of c] (d);
						\vertex [above left=of a] (o1);
						\vertex [below left=of a] (o2);
						\vertex [above=of b] (be);
						\vertex [above=of c] (ce);
						\diagram* {
							(a) -- [anti fermion] (b) -- [fermion] (c) -- [fermion] (d),
							(b) -- [photon] (be),
							(c) -- [photon] (ce),
							(a) -- [photon] (o1),
							(a) -- [photon] (o2),
						};
					\end{feynman}\;
			\end{tikzpicture}
			\ +\ (1{+}\theta^-)
			\begin{tikzpicture}[baseline={([yshift=-.16cm]current bounding box.center)}, scale=0.4, transform shape]
					\begin{feynman}
						\vertex (a);
						\vertex [right=of a] (b);
						\vertex [right=of b] (c);
						\vertex [right=of c] (d);
						\vertex [above left=of a] (o1);
						\vertex [below left=of a] (o2);
						\vertex [above=of b] (be);
						\vertex [above=of c] (ce);
						\diagram* {
							(a) -- [anti fermion] (b) -- [anti fermion] (c) -- [fermion] (d),
							(b) -- [photon] (be),
							(c) -- [photon] (ce),
							(a) -- [photon] (o1),
							(a) -- [photon] (o2),
						};
					\end{feynman}\;
			\end{tikzpicture}
			\ +\ \theta^+
			\begin{tikzpicture}[baseline={([yshift=-.16cm]current bounding box.center)}, scale=0.4, transform shape]
					\begin{feynman}
						\vertex (a);
						\vertex [right=of a] (b);
						\vertex [right=of b] (c);
						\vertex [right=of c] (d);
						\vertex [above left=of a] (o1);
						\vertex [below left=of a] (o2);
						\vertex [above=of b] (be);
						\vertex [above=of c] (ce);
						\diagram* {
							(a) -- [fermion] (b) -- [anti fermion] (c) -- [fermion] (d),
							(b) -- [photon] (be),
							(c) -- [photon] (ce),
							(a) -- [photon] (o1),
							(a) -- [photon] (o2),
						};
					\end{feynman}\;
			\end{tikzpicture}\\
			&
			\ +\ 2(1{+}\theta^+)
			\begin{tikzpicture}[baseline={([yshift=-.4cm]current bounding box.center)}, scale=0.4, transform shape]
					\begin{feynman}
						\vertex (a);
						\vertex [right=of a] (b);
						\vertex [right=of b] (c);
						\vertex [above left=of a] (d);
						\vertex [below left=of a] (e);
						\vertex [above=of b] (f);
						\vertex [above right=of f] (f1);
						\vertex [above left=of f] (f2);
						\diagram* {
							(a) -- [fermion] (b) -- [fermion] (c),
							(b) -- [fermion] (f),
							(f) -- [photon] (f1),
							(f) -- [photon] (f2),
							(a) -- [photon] (d),
							(a) -- [photon] (e),
						};
					\end{feynman}\;
			\end{tikzpicture}
			\ +\ \theta^- \begin{tikzpicture}[baseline={([yshift=-.4cm]current bounding box.center)}, scale=0.4, transform shape]
					\begin{feynman}
						\vertex (a);
						\vertex [right=of a] (b);
						\vertex [right=of b] (c);
						\vertex [above left=of a] (d);
						\vertex [below left=of a] (e);
						\vertex [above=of b] (f);
						\vertex [above right=of f] (f1);
						\vertex [above left=of f] (f2);
						\diagram* {
							(a) -- [anti fermion] (b) -- [fermion] (c),
							(b) -- [fermion] (f),
							(f) -- [photon] (f1),
							(f) -- [photon] (f2),
							(a) -- [photon] (d),
							(a) -- [photon] (e),
						};
					\end{feynman}\;
			\end{tikzpicture}\Bigr]
			\ +\ \bigl(\theta^+\leftrightarrow\theta^-,\Sr\leftrightarrow\Sl\bigr)\Bigr\}\ +\ \mathcal{O}(g^4)\ .
		\end{aligned}
	\end{equation}
	It appears to be difficult to derive a formula for the $\theta$~polynomial multiplying a general tree.
	We now turn to computing bosonic correlation functions using the characteristic property \eqref{eq:globalflow} of the Nicolai map.
	
	\section{Amplitudes in supersymmetric quantum mechanics}\label{sec:sqm_amplitudes}
	In this section, we compute the one-, two-, and three-point function for the interacting SQM theory from \ref{subsec:interacting}. 
	In order to compute the loop integrals, we use analytic continuation and Wick rotate to Euclidean space ($\omega\rightarrow \im\omega$), 
	so that the free Euclidean propagators become
	\begin{equation}\label{eq:E_fermion_rules}
	S_0(\omega)\quad\=\quad\begin{tikzpicture}[baseline={([yshift=-.1cm]current bounding box.center)}]
		\begin{feynman}
			\vertex (a);
			\vertex [right=of a] (b);
			\diagram* {
				(a) -- [fermion] (b),
			};
		\end{feynman}\;\end{tikzpicture}\quad \=\quad\frac{\im\omega-m}{\omega^2+m^2}\ ,\qquad
	S_0(-\omega)\quad\=\quad\begin{tikzpicture}[baseline={([yshift=-.1cm]current bounding box.center)}]
		\begin{feynman}
			\vertex (a);
			\vertex [right=of a] (b);
			\diagram* {
				(a) -- [anti fermion] (b),
			};
		\end{feynman}\;\end{tikzpicture}\quad \=\quad\frac{-\im\omega-m}{\omega^2+m^2}\ ,
	\end{equation}
	and
	\begin{equation}
	G_0(\omega)\quad=\quad\begin{tikzpicture}[baseline={([yshift=-.1cm]current bounding box.center)}]
		\begin{feynman}
			\vertex (a);
			\vertex [right=of a] (b);
			\diagram* {
				(a) -- [photon] (b),
			};
		\end{feynman}\;\end{tikzpicture}\quad \=\quad\frac{-1}{\omega^2+m^2}\ .
	\end{equation}

	\noindent\textbf{One-point function.\ }
	We need to contract the open boson lines of \eqref{eq:inv_map_interacting_theta}. At one-loop order, this gives two diagrams,
	\begin{equation}
		\begin{aligned}
			\bigl\langle x(\omega)\bigr\rangle_g\=&g\theta^+\  \begin{tikzpicture}[baseline={([yshift=-.1cm]current bounding box.center)}]
				\begin{feynman}
					\vertex (b) at ($(0,0)$);
					\vertex (s) at ($(b) + (0, .7cm)$);
					\draw[fermion]  (s) -- (b);
					\draw[photon] (s) arc [start angle=-90, end angle=270, radius=0.4cm];
				\end{feynman}
			\end{tikzpicture}
			\ +\ 
			g\theta^-\  \begin{tikzpicture}[baseline={([yshift=-.1cm]current bounding box.center)}]
				\begin{feynman}
					\vertex (b) at ($(0,0)$);
					\vertex (s) at ($(b) + (0, .7cm)$);
					\draw[fermion]  (b) -- (s);
					\draw[photon] (s) arc [start angle=-90, end angle=270, radius=0.4cm];
				\end{feynman}
			\end{tikzpicture}\ +\ \mathcal{O}(g^3)\\ \=&2\pi\delta(\omega)\;g\,(\theta^+{+}\theta^-)\;\sfrac{-m}{m^2}\int\sfrac{\diff l}{2\pi}\sfrac{-1}{l^2+m^2}\ +\ \mathcal{O}(g^3)\=g\,\sfrac{\pi\delta(\omega)}{m^2}\ +\ \mathcal{O}(g^3)\ ,
		\end{aligned}
	\end{equation}
	where we used the loop integral
	\begin{equation}
		\int_{l}G_0(l)\=\int\sfrac{\diff l}{2\pi}\ \sfrac{-1}{m^2+l^2}\=-\sfrac{1}{2m}\ .
	\end{equation}
	The final result is independent of $\theta$ as expected. 
	It further matches the expression \eqref{eq:feyn_1point} from conventional Feynman perturbation theory, see Appendix~\ref{app:feyn}.
	
	\noindent\textbf{Two-point function.\ } 
	For the two-point function we need to multiply the inverse map expansion with itself and then contract the open boson lines. We find the following (connected) diagrams:
	\begin{equation}
		\begin{aligned}
			\bigl\langle x(\omega)\ x(\omega')\bigr\rangle_g\=&2\pi\delta(\omega{+}\omega')\;G_0(\omega)\\
			+&\ 2g^2\Bigl\{{\theta^+}^2\begin{tikzpicture}[baseline={([yshift=-.1cm]current bounding box.center)},scale=0.5, transform shape]
					\begin{feynman}
						\vertex (a);
						\vertex [right=of a] (b);
						\vertex [right=of b] (c);
						\vertex [right=of c] (d);
						\diagram*{
							(a) -- [anti fermion] (b),
							(b) -- [photon, half left, looseness=1.5] (c),
							(c) -- [photon, half left, looseness=1.5] (b),
							(c) -- [fermion] (d),
						};\end{feynman}
					\end{tikzpicture}\ +\ \theta^+\theta^-\begin{tikzpicture}[baseline={([yshift=-.1cm]current bounding box.center)},scale=0.5, transform shape]
					\begin{feynman}
						\vertex (a);
						\vertex [right=of a] (b);
						\vertex [right=of b] (c);
						\vertex [right=of c] (d);
						\diagram*{
							(a) -- [fermion] (b),
							(b) -- [photon, half left, looseness=1.5] (c),
							(c) -- [photon, half left, looseness=1.5] (b),
							(c) -- [fermion] (d),
						};\end{feynman}
				\end{tikzpicture}\ +\ \bigl(\theta^+\leftrightarrow\theta^-,\Sr\leftrightarrow\Sl\bigr)\Bigr\}\\
		+&\ \sfrac{g^2}{2}\Bigl\{\theta^+(1{+}\theta^+)(2\  \begin{tikzpicture}[baseline={([yshift=-.1cm]current bounding box.center)},scale=0.5, transform shape]
				\begin{feynman}
					\vertex (a);
					\vertex [right=of a] (b);
					\vertex [right=of b] (c);
					\vertex [right=of c] (d);
					\diagram*{
						(a) -- [photon] (b),
						(b) -- [photon, half left, looseness=1.5] (c),
						(c) -- [anti fermion, half left, looseness=1.5] (b),
						(c) -- [fermion] (d),
					};\end{feynman}
		\end{tikzpicture}\ +\ \begin{tikzpicture}[baseline={([yshift=-.65cm]current bounding box.center)},scale=0.7, transform shape]
			\begin{feynman}
				\vertex (a);
				\vertex [right=of a] (b);
				\vertex [right=of b] (c);
				\vertex (s) at ($(b) + (0, .7cm)$);
				\draw  (b)[anti fermion] -- (s);
				\draw (s) [photon] arc [start angle=-90, end angle=270, radius=0.5cm];
				\diagram*{
					(a) -- [photon] (b) -- [fermion] (c),
				};
			\end{feynman}
		\end{tikzpicture})\\
		&\hspace{1.2cm} +\ \theta^+\theta^-(2\  \begin{tikzpicture}[baseline={([yshift=-.1cm]current bounding box.center)},scale=0.5, transform shape]
				\begin{feynman}
					\vertex (a);
					\vertex [right=of a] (b);
					\vertex [right=of b] (c);
					\vertex [right=of c] (d);
					\diagram*{
						(a) -- [photon] (b),
						(b) -- [photon, half left, looseness=1.5] (c),
						(c) -- [anti fermion, half left, looseness=1.5] (b),
						(c) -- [anti fermion] (d),
					};\end{feynman}
		\end{tikzpicture}\ +\ \begin{tikzpicture}[baseline={([yshift=-.65cm]current bounding box.center)},scale=0.7, transform shape]
		\begin{feynman}
			\vertex (a);
			\vertex [right=of a] (b);
			\vertex [right=of b] (c);
			\vertex (s) at ($(b) + (0, .7cm)$);
			\draw  (b)[fermion] -- (s);
			\draw (s) [photon] arc [start angle=-90, end angle=270, radius=0.5cm];
			\diagram*{
				(a) -- [photon] (b) -- [fermion] (c),
			};
		\end{feynman}
	\end{tikzpicture})\ +\ \bigl(\theta^+\leftrightarrow\theta^-,\Sr\leftrightarrow\Sl\bigr)\Bigr\}\\
		+&\ \sfrac{g^2}{2}\Bigl\{\omega\leftrightarrow\omega'\Bigr\}\ +\ \mathcal{O}(g^4)\ .
		\end{aligned}
	\end{equation}
	With the loop integrals
	\begin{equation}
		\int_{l}G_0(l)G_0(\omega-l)\=\int\sfrac{\diff l}{2\pi}\ \sfrac{-1}{m^2+l^2}\sfrac{-1}{m^2+(\omega-l)^2}\=\sfrac{1}{4m^3+m\omega^2}\ ,
	\end{equation}
	\begin{equation}
		\int_{l}G_0(l)S_0(\omega+l)\=\int\sfrac{\diff l}{2\pi}\ \sfrac{-1}{m^2+l^2}\sfrac{\im(\omega+l)-m}{m^2+(\omega+l)^2}\=\sfrac12 \sfrac{2m-\im\omega}{4m^3+m\omega^2}
	\end{equation}
	we find
	\begin{equation}
		\begin{aligned}
			\bigl\langle x(\omega)\ x(\omega')\bigr\rangle_g\ =&\ 2\pi\delta(\omega{+}\omega')\;G_0(\omega)\\ +&\ 2g^2 \sfrac{2\pi\delta(\omega+\omega')}{(\omega^2+m^2)^2}\Bigl\{\sfrac{{\theta^+}^2(\im\omega'-m)(\im\omega-m)\ +\ \theta^+\theta^-(-\im\omega'-m)(\im\omega-m)}{4m^3+m\omega^2}\ +\ \bigl(\theta^+\leftrightarrow\theta^-,\omega\leftrightarrow\omega'\bigr)\Bigr\}\\
			+&\ \sfrac{g^2}{2}\sfrac{2\pi\delta(\omega+\omega')}{(\omega^2+m^2)^2}\Bigl\{\theta^+(1{+}\theta^+)(-4\ \sfrac12 \sfrac{(\im\omega-m)(2m-\im\omega)}{4m^3+m\omega^2}\ -\ 2 \sfrac{\im\omega-m}{2m^2})\\
			&\hspace{2.5cm} +\ \theta^+\theta^-(-4\ \sfrac12 \sfrac{(-\im\omega-m)(2m-\im\omega)}{4m^3+m\omega^2}\ -\ 2 \sfrac{\im\omega-m}{2m^2})\ +\ \bigl(\theta^+\leftrightarrow\theta^-,\omega\leftrightarrow\omega'\bigr)\Bigr\}\\
			+&\ \sfrac{g^2}{2}\sfrac{2\pi\delta(\omega+\omega')}{(\omega^2+m^2)^2}\Bigl\{\omega\rightarrow -\omega\Bigr\}\ +\ \mathcal{O}(g^4)\ .
		\end{aligned}
	\end{equation}
	Simplifying this yields
	\begin{equation}
		\begin{aligned}
			\bigl\langle x(\omega)\ x(\omega')\bigr\rangle_g\=&2\pi\delta(\omega+\omega')G_0(\omega)\ +\ 2g^2 \sfrac{2\pi\delta(\omega+\omega')}{(\omega^2+m^2)^2}\sfrac{({\theta^+}^2+{\theta^-}^2) (m^2+\omega^2)\ +\ 2\theta^+\theta^-(m^2-\omega^2)}{4m^3+m\omega^2}\\
			+&\ \sfrac{g^2}{2}\sfrac{2\pi\delta(\omega+\omega')}{(\omega^2+m^2)^2}\Bigl\{[1+{\theta^+}^2+{\theta^-}^2](4\  \sfrac{2m^2-\omega^2}{4m^3+m\omega^2}\ +\  \sfrac{2}{m})\\
			&\hspace{2.85cm} +\ 2\theta^+\theta^-(4\  \sfrac{2m^2+\omega^2}{4m^3+m\omega^2}\ +\  \sfrac{2}{m})\Bigr\}
			\ +\ \mathcal{O}(g^4)\ ,
		\end{aligned}
	\end{equation}
	and finally (using $\theta^++\theta^-=1$ and $\theta^+-\theta^-=\theta$)
	\begin{equation}
		\bigl\langle x(\omega)\ x(\omega')\bigr\rangle_g\=2\pi\delta(\omega+\omega')G_0(\omega)\ +\ g^2\ \sfrac{2\pi\delta(\omega+\omega')}{(m^2+\omega^2)^2}\sfrac{18m^2}{4m^3+m\omega^2}\ +\ \mathcal{O}(g^4)\ ,
	\end{equation}
	again matching the expression obtained from Feynman diagrams \eqref{eq:feyn_2point}.
	
	\noindent\textbf{Three-point function.\ } 
	For simplicity, here we restrict ourselves to the cases $\theta=\pm 1$, 
	so we can use \eqref{eq:inv_map_interacting_pm} as the inverse map with the fermion propagator $S_0(\pm\omega)$ \eqref{eq:E_fermion_rules} for $\theta=\pm 1$.
	For the three-point function, one needs to determine
	\begin{equation}
		\bigl\langle x(\omega_1)\ x(\omega_2)\ x(\omega_3)\bigr\rangle_g\=
		\bigl\langle T_g^{-1} x(\omega_1)\ T_g^{-1} x(\omega_2)\ T_g^{-1} x(\omega_3)\bigr\rangle_0\ .
	\end{equation}
	At order $g^3$ one needs to evaluate (free) correlators of six bosonic fields. 
	Using Wick's theorem, each of them gives 15 diagrams, of which some are disconnected, which we ignore, and many others are equivalent. 
	Naturally, everything will be totally symmetric in $(\omega_1,\omega_2,\omega_3)$. 
	Up to permutations of the frequencies, there are 11 diagrams, of which 4 are one-particle irreducible (1PI). We label the latter four as
	\begin{equation}
		\begin{aligned}
			&N_1\=
			\begin{tikzpicture}[baseline={([yshift=-.1cm]current bounding box.center)}]
				\begin{feynman}
					\vertex (a1) at ($(0,0)$);
					\vertex (a2) at ($(1,0)$);
					\vertex (a3) at ($(.5,-.87)$);
					\vertex [above left=1cm of a1] (e1);
					\vertex [above right=1cm of a2] (e2);
					\vertex [below =1cm of a3] (e3);
					\diagram*{
						(a1) -- [photon] (a2) -- [photon] (a3) -- [photon] (a1),
						(a1) -- [] (e1),
						(a2) -- [] (e2),
						(a3) -- [] (e3),
					};\end{feynman}
			\end{tikzpicture}
			\ ,\qquad 
			N_2\=\begin{tikzpicture}[baseline={([yshift=-.1cm]current bounding box.center)}]
				\begin{feynman}
					\vertex (a1) at ($(0,0)$);
					\vertex (a2) at ($(1,0)$);
					\vertex (a3) at ($(.5,-.87)$);
					\vertex [above left=1cm of a1] (e1);
					\vertex [above right=1cm of a2] (e2);
					\vertex [below =1cm of a3] (e3);
					\diagram*{
						(a1) -- [] (a2) -- [photon] (a3) -- [photon] (a1),
						(a1) -- [photon] (e1),
						(a2) -- [] (e2),
						(a3) -- [] (e3),
					};\end{feynman}
			\end{tikzpicture}\ ,\\
			&N_3\=\begin{tikzpicture}[baseline={([yshift=-.1cm]current bounding box.center)}]
				\begin{feynman}
					\vertex (a1) at ($(0,0)$);
					\vertex (a2) at ($(1,0)$);
					\vertex (a3) at ($(.5,-.87)$);
					\vertex [above left=1cm of a1] (e1);
					\vertex [above right=1cm of a2] (e2);
					\vertex [below =1cm of a3] (e3);
					\diagram*{
						(a1) -- [photon] (a2) -- [] (a3) -- [] (a1),
						(a1) -- [photon] (e1),
						(a2) -- [] (e2),
						(a3) -- [photon] (e3),
					};\end{feynman}
			\end{tikzpicture}\ ,\qquad
			N_4\=\begin{tikzpicture}[baseline={([yshift=-.1cm]current bounding box.center)}]
				\begin{feynman}
					\vertex (a1) at ($(0,0)$);
					\vertex (a2) at ($(1,0)$);
					\vertex (a3) at ($(.5,-.87)$);
					\vertex [above left=1cm of a1] (e1);
					\vertex [above right=1cm of a2] (e2);
					\vertex [below =1cm of a3] (e3);
					\diagram*{
						(a1) -- [photon] (a2) -- [] (a3) -- [] (a1),
						(a1) -- [photon] (e1),
						(a2) -- [photon] (e2),
						(a3) -- [] (e3),
					};\end{feynman}
			\end{tikzpicture}\ .
		\end{aligned}
	\end{equation}
	Furthermore, there are five one-particle reducible (1PR) diagrams that combine to give the 2-point function on one leg. We collect them under the label
	\begin{equation}
		N_5\=
		\begin{tikzpicture}[baseline={([yshift=-.1cm]current bounding box.center)}]
			\begin{feynman}
				\vertex (b);
				\vertex [above left=of b](a1);
				\vertex [below left=of b](a2);
				\vertex [right =1cm of b] (c) [blob,label=\(\textrm{2-point}\)] {};
				\vertex [right =of c] (d);
				\diagram*{
					(a1) -- [photon] (b),
					(a2) -- [] (b),
					(b) -- [photon] (c) -- [photon] (d),
				};\end{feynman}
		\end{tikzpicture}
		\ .
	\end{equation}
	Finally, there are two more reducible diagrams,
	\begin{equation}
		N_6\=
		\begin{tikzpicture}[baseline={([yshift=-.1cm]current bounding box.center)},scale=.8,transform shape]
			\begin{feynman}
				\vertex (b);
				\vertex [above left=of b](a1);
				\vertex [below left=of b](a2);
				\vertex [right = of b] (c);
				\vertex [right =of c] (d);
				\vertex [right =of d] (e);
				\diagram*{
					(a1) -- [photon] (b),
					(a2) -- [photon] (b),
					(b) -- (c) -- [photon, half left, looseness=1.5] (d)
					-- [half left, looseness=1.5] (c),
					(d) -- (e),
				};\end{feynman}
		\end{tikzpicture}
		\ ,\qquad
		N_7\=\begin{tikzpicture}[baseline={([yshift=-.4cm]current bounding box.center)},scale=.8,transform shape]
			\begin{feynman}
				\vertex (a);
				\vertex [right =of a] (b);
				\vertex [right =of b] (c);
				\vertex (s) at ($(b) + (0, .7cm)$);
				\vertex [above left=of a] (o1);
				\vertex [below left=of a] (o2);
				\draw (o1) [photon] -- (a);
				\draw (o2) [photon] -- (a);
				\draw (b) -- (s);
				\draw (s) [photon] arc [start angle=-90, end angle=270, radius=0.5cm];
				\diagram* {
					(a) -- [] (b) -- [] (c),
				};
			\end{feynman}
		\end{tikzpicture}\ .
	\end{equation}
	To represent the various contributions, we pull out an overall factor of
	\begin{equation}\label{eq:prefactor}
		g^3\ 2\pi\delta(\omega_1{+}\omega_2{+}\omega_3)\prod_{i=1,2,3} (m^2+\omega_i^2)^{-2} (4m^2+\omega_i^2)^{-1}
	\end{equation}
	and express the remainder of all contributions in terms of the symmetric polynomials
	\begin{equation}\label{eq:sym_pol}
		t_1\=\omega_1+\omega_2+\omega_3\ \equiv\ 0\ ,\qquad t_2\=\omega_1\omega_2+\omega_1\omega_3+\omega_2\omega_3\ ,\qquad
		t_3\=\omega_1\omega_2\omega_3\ .
	\end{equation}
	We find for the 1PI contributions:
	\begin{equation}
		\begin{aligned}
			N_1\quad\rightarrow\quad &96 m^{10} - 296 m^8 t_2 + 312 m^6 t_2^2 - 120 m^4 t_2^3 + 8 m^2 t_2^4 \pm 
			96 \im m^7 t_3 \mp 200 \im m^5 t_2 t_3 \pm 112 \im m^3 t_2^2 t_3 \mp 8 \im m t_2^3 t_3  \\ + &
			96 m^4 t_3^2 - 104 m^2 t_2 t_3^2 + 8 t_2^2 t_3^2 \pm 96 \im m t_3^3 \mp 
			8 \im m^{-1} t_2 t_3^3\ ,\\
			N_2\quad\rightarrow\quad &576 m^{10} - 1200 m^8 t_2 + 672 m^6 t_2^2 - 48 m^4 t_2^3 \mp 288 \im m^7 t_3 \pm 
			600 \im m^5 t_2 t_3 \mp 336 \im m^3 t_2^2 t_3 \pm 24 \im m t_2^3 t_3\\ + &576 m^4 t_3^2 - 48 m^2 t_2 t_3^2 \mp 288 \im m t_3^3 \pm 24 \im m^{-1} t_2 t_3^3\ ,
			&\\
			N_3\quad\rightarrow\quad &384 m^{10} - 608 m^8 t_2 + 48 m^6 t_2^2 + 192 m^4 t_2^3 - 16 m^2 t_2^4 \pm 
			96 \im m^7 t_3 \mp 200 \im m^5 t_2 t_3 \pm 112 \im m^3 t_2^2 t_3 \mp 8 \im m t_2^3 t_3 \\ + 
			&384 m^4 t_3^2 + 160 m^2 t_2 t_3^2 - 16 t_2^2 t_3^2 \pm 96 \im m t_3^3 \mp 
			8 \im m^{-1} t_2 t_3^3\ ,\\
			N_4\quad=\quad &N_3\ .
		\end{aligned}
	\end{equation}
	Summing these up, the ($\theta$-dependent) imaginary terms drop out, and we are left with
	\begin{equation}
		\begin{aligned}
			N_{\mathrm{1PI}}\=&N_1+N_2+N_3+N_4\\ \quad\rightarrow\quad &1440 m^{10} - 2712 m^8 t_2 + 1080 m^6 t_2^2 + 216 m^4 t_2^3 - 
			24 m^2 t_2^4 + 1440 m^4 t_3^2 + 168 m^2 t_2 t_3^2 - 24 t_2^2 t_3^2\ .
		\end{aligned}
	\end{equation}
	Similarly, the reducible contributions yield:
	\begin{equation}
		\begin{aligned}
			N_5\quad\rightarrow\quad &3456 m^{10} - 5760 m^8 t_2 + 2952 m^6 t_2^2 - 720 m^4 t_2^3 + 
			72 m^2 t_2^4 \mp 2160 \im m^7 t_3 \pm 1620 \im m^5 t_2 t_3 \\ \mp &360 \im m^3 t_2^2 t_3 \pm 
			36 \im m t_2^3 t_3 - 1080 m^4 t_3^2 + 288 m^2 t_2 t_3^2 \pm 108 \im m t_3^3\ ,\\
			N_6\quad\rightarrow\quad &384 m^{10} - 128 m^8 t_2 - 312 m^6 t_2^2 + 48 m^4 t_2^3 + 8 m^2 t_2^4 \pm 
			1392 \im m^7 t_3 \mp 980 \im m^5 t_2 t_3 \\ \pm &184 \im m^3 t_2^2 t_3 \mp 
			20 \im m t_2^3 t_3 - 1128 m^4 t_3^2 + 352 m^2 t_2 t_3^2 - 16 t_2^2 t_3^2 \mp 
			120 \im m t_3^3 \pm 4 \im m^{-1} t_2 t_3^3\ ,
			&\\
			N_7\quad\rightarrow\quad &384 m^{10} - 448 m^8 t_2 + 24 m^6 t_2^2 + 48 m^4 t_2^3 - 8 m^2 t_2^4 \pm 
			768 \im m^7 t_3 \mp 640 \im m^5 t_2 t_3 \\ \pm &176 \im m^3 t_2^2 t_3 \mp 
			16 \im m t_2^3 t_3 - 378 m^4 t_3^2 + 188 m^2 t_2 t_3^2 - 26 t_2^2 t_3^2 \pm 
			12 \im m t_3^3 \mp 4 \im m^{-1} t_2 t_3^3 - 6 m^{-2} t_3^4\ ,
		\end{aligned}
	\end{equation}
	which adds up to
	\begin{equation}
		\begin{aligned}
			&N_{\mathrm{1PR}}\=N_5+N_6+N_7\\ &\rightarrow\quad 4224 m^{10} - 6336 m^8 t_2 + 2664 m^6 t_2^2 - 624 m^4 t_2^3 + 
			72 m^2 t_2^4 - 2586 m^4 t_3^2 + 828 m^2 t_2 t_3^2 - 42 t_2^2 t_3^2 - 
			6 m^{-2} t_3^4\ .
		\end{aligned}
	\end{equation}
	The final result (multiplying the prefactor~\eqref{eq:prefactor}) reads 
	\begin{equation}\label{eq:N_final}
		\begin{aligned}
			&N\=N_{\mathrm{1PI}}+N_{\mathrm{1PR}}\\ &\rightarrow\quad 5664 m^{10} - 9048 m^8 t_2 + 3744 m^6 t_2^2 - 408 m^4 t_2^3 + 
			48 m^2 t_2^4 - 1146 m^4 t_3^2 + 996 m^2 t_2 t_3^2 - 66 t_2^2 t_3^2 - 
			6 m^{-2} t_3^4\ .
		\end{aligned}
	\end{equation}
	Consulting Appendix~\ref{app:feyn} we confirm perfect agreement with the result~\eqref{eq:feyn_3point} from Feynman diagrams.
	
	While these calculations are quite technical, we can make some interesting observations, in particular from the computation of the three-point function. 
	It is curious that the irreducible contributions $N_{\mathrm{1PI}}$ and the reducible contributions $N_{\mathrm{1PR}}$ are $\theta$-independent separately.\footnote{
	    The same holds for the two-point function.} 
	Moreover, the $\theta$-dependent terms that appear in the single diagrams are all imaginary and all proportional to an odd power of $t_3=\omega_1\omega_2\omega_3$.
	Finally, the notion of one-particle irreducibility differs from that for Feynman diagrams. 
	The diagrams $N_6$ and $N_7$ naively appear to be 1PR, but they do not decompose into 1PI subdiagrams, 
	because the Nicolai rules do not allow for fermion lines connecting external points of a diagram.
	For Nicolai diagrams the reducibility notion thus is restricted to cutting boson lines only, whence $N_6$ and $N_7$ should be counted as 1PI as well.
    
	\bigskip 

	\noindent{\bf\large Acknowledgments.\ }\\
	M.R.~is supported by a PhD grant of the German Academic Scholarship Foundation.

	\appendix

	\section{Nicolai map for a multi-coupling straight contour}\label{app:straight}
	Here, we give an explicit perturbative expansion for the multi-coupling Nicolai map
	\begin{equation}
        T_g[h]\;\phi 
        \= \sum_{n=0}^{\infty}(-)^n\int_{0}^{1}\drm s_n\int_0^{s_n}\drm s_{n-1}\cdots\int_0^{s_2}\drm s_1\ 
        \Bigl[\vec{h}'(s_n)\cdot \vec{R}\bigl(h(s_n)\bigr)\Bigr]\cdots\Bigl[\vec{h}'(s_1)\cdot \vec{R}\bigl(h(s_1)\bigr)\Bigr]\ \phi
        \end{equation}
	in the case of the straight contour, $h=\scontour$,
	\begin{equation}
	\vec{h}(s) \= s\;\vec{g} \qquad\textrm{for}\quad \vec{g}\=\bigl(g_1,g_2,\ldots,g_k\bigr)\ .
	\end{equation}
	We start by expanding
        \begin{equation}
        g_i\,R^{(i)}(s\,g)\=\sum_{\alpha} s^{|\alpha|-1}g^{\alpha}\,\rR^{(i)}_\alpha\ ,\qquad (\text{no sum over }i)
        \end{equation}
        where $\alpha$ is a multi-index $(\alpha_1,\ldots,\alpha_k)$ with $\alpha_i\geq 1$ and $\alpha_l\geq 0$ for $l{\neq}i$.
        Using $n$ such multi-indices $\alpha^1,\ldots,\alpha^n$, we can express the Nicolai map as
        \begin{equation}
        T_g[\scontour]\;\phi
        \= \sum_{n=0}^{\infty}\ \sum_{i_1,\ldots, i_n=1}^k\  \sum_{\alpha^1,\ldots,\alpha^n}\!\!(-)^n
        \int_{0}^{1}\!\!\!\drm s_n\int_0^{s_n}\!\!\!\drm s_{n-1}\cdots\int_0^{s_2}\!\!\!\drm s_1\ s_n^{|\alpha^n|-1}\ldots s_1^{|\alpha^1|-1}g^{\alpha^n}\cdots g^{\alpha^1}
        \rR^{(i_n)}_{\alpha^n}\cdots \rR^{(i_1)}_{\alpha^1} \phi\ .
        \end{equation}
        Carrying out the integrals, one obtains the coefficients
        \begin{equation}
        \begin{aligned}
        c_\alpha\ :=\ &(-)^n\int_{0}^{1}\!\!\drm s_n\;s_n^{|\alpha^n|-1} \cdots \int_0^{s_3}\!\!\drm s_{2}\;s_2^{|\alpha^2|-1}\int_0^{s_2}\!\!\drm s_{1}\;s_1^{|\alpha^1|-1}\\ 
        \=&(-)^n\Bigl[|\alpha^1|\cdot\bigl(|\alpha^1|+|\alpha^2|\bigr)\cdots\bigl(|\alpha^1|+\ldots+|\alpha^n|\bigr)\Bigr]^{-1}\ ,
        \end{aligned}
        \end{equation}
        a natural generalization of the coefficients for the map with only one coupling \cite{LR1}. The final formula can be written compactly as
        \begin{equation}
        T_g[\scontour]\;\phi
                \= \sum_{n=0}^{\infty}\ \sum_{i_1,\ldots, i_n=1}^k \sum_{\alpha}\ c_\alpha\ g^\alpha\ 
                \rR^{(i_n)}_{\alpha^n}\cdots \rR^{(i_1)}_{\alpha^1}\ \phi\ ,
        \end{equation}
        where $\alpha=(\alpha^1,\ldots,\alpha^n)$ is a multi-index of $n$ multi-indices ($i=1,\ldots,n$)
        \begin{equation}
        \alpha^i\=\bigl((\alpha^i)_1,\ldots,(\alpha^i)_k\bigr)
        \end{equation}
    	that takes values
        \begin{equation}
        (\alpha^p)_{i_q}\ \geq\ 0\quad \text{for}\quad p\ \neq\ q\ ,\qquad\qquad (\alpha^p)_{i_p}\ \geq\ 1\ ,
        \end{equation}
        and
        \begin{equation}
        g^{\alpha}\=g^{\alpha^1}\cdots\ g^{\alpha^n}\=g_1^{\sum_{l=1}^{n}(\alpha^l)_1}\cdots\ g_k^{\sum_{l=1}^{n}(\alpha^l)_k}\ .
        \end{equation}

	\section{Proof of free-action and determinant-matching property}\label{app:n_proofs}
	\noindent\textbf{Free action.\ } Here we check whether our flow operators satisfy their respective infinitesimal free-action conditions
	\begin{equation}
		(\partial_g + R_g)\ S^{\mathrm{b}}\=0\ ,\qquad (\partial_\theta + R_\theta)\ S^{\mathrm{b}}\=0\ ,
	\end{equation}
	for the bosonic action
	\begin{equation}
		S^{\mathrm{b}}\=\int \diff t\ \bigl\{\sfrac{1}{2}\dot{x}_i^2-\sfrac12 V_i(x)^2+\im\theta \sfrac{\diff}{\diff t}V(x(t))\bigr\}\ .
	\end{equation}
	Beginning with the $g$~flow, we have
	\begin{equation}
		R_g^+ S^{\mathrm{b}}\=-\int\diff t\ \diff t'\ \partial_g V_i(t) S_{ij}(t',t)(\ddot{x}_j+V_{jk}V_k)(t')\ ,
	\end{equation}
	and $R_g^- S^{\mathrm{b}}$ with the reversed arguments in $S_{ij}$.
	Now we use \eqref{eq:ferm_prop} to replace
	\begin{equation}\label{eq:replacements}
		S_{ij}(t',t)V_{jk}(t')\ \longrightarrow\ -\im S_{ik}(t',t)\diff_{t'}-\delta_{ik}\delta(t-t')\ ,\qquad 
		S_{ij}(t,t')V_{jk}(t')\ \longrightarrow\ +\im S_{ik}(t,t')\diff_{t'}-\delta_{ik}\delta(t-t')
	\end{equation}
	under the integrals. Doing this once leads to
	\begin{equation}
		\begin{aligned}
			&R_g^+ S^{\mathrm{b}}\=-\int\diff t\ \diff t'\ \partial_g V_i(t) S_{ij}(t',t)\ddot{x}_j(t')+\im\int\diff t\ \diff t'\ \partial_g V_i(t) S_{ij}(t',t)V_{jk}(t')\dot{x}_k(t')
			+\int\diff t\ (\partial_g V_i) V_i\ ,\\
			&R_g^- S^{\mathrm{b}}\=-\int\diff t\ \diff t'\ \partial_g V_i(t) S_{ij}(t,t')\ddot{x}_j(t')+\im\int\diff t\ \diff t'\ \partial_g V_i(t) S_{ij}(t,t')V_{jk}(t')\dot{x}_k(t')+\int\diff t\ (\partial_g V_i) V_i\ ,
		\end{aligned}
	\end{equation}
	and doing it once more cancels most terms, leaving us with
	\begin{equation}
		\begin{aligned}
			&R_g^+ S^{\mathrm{b}}\=-\im\int\diff t\ (\partial_g V_i)\; \dot{x}_i+\int\diff t\ (\partial_g V_i) V_i\ ,\\
			&R_g^- S^{\mathrm{b}}\=+\im\int\diff t\ (\partial_g V_i)\; \dot{x}_i+\int\diff t\ (\partial_g V_i) V_i\ .
		\end{aligned}
	\end{equation}
	We conclude that as expected,
	\begin{equation}
		(\partial_g + R_g)S^{\mathrm{b}}\=(\partial_g + \theta^+R_g^++\theta^- R_g^-)S^{\mathrm{b}}\=\int \diff t\ \bigl[-V_i\partial_gV_i+\im\theta \sfrac{\diff}{\diff t}(\partial_g V)\bigr]+\int \diff t\ \bigl[V_i\partial_gV_i-\im\theta \sfrac{\diff}{\diff t}(\partial_g V)\bigr]\=0\ .
	\end{equation}
	Let us proceed with the $\theta$~flow in the form \eqref{eq:theta_flow}. From a similar calculation, we find
	\begin{equation}
		\begin{aligned}
			R_\theta^\beta S^{\mathrm{b}}\=-\sfrac{\beta}{2} \int\diff t\ \diff t'\ V_i(t) &[S_{ij}(t',t)-S_{ij}(t,t')](\ddot{x}_j+V_{jk}V_k)(t')\\
			-\ \im\ \sfrac{\beta-1}{2}\int\diff t\ \diff t'\  
			\dot{x}_i(t) &[S_{ij}(t',t)+S_{ij}(t,t')](\ddot{x}_j+V_{jk}V_k)(t')\ .
		\end{aligned}
	\end{equation}
	Again we can insert the replacements \eqref{eq:replacements} twice and end up with
	\begin{equation}
			R_\theta^\beta S^{\mathrm{b}}\=-\im\beta \int\diff t\ V_i\dot{x}_i\ +\ \im(\beta{-}1)\int\diff t\ \dot{x}_iV_i\=-\im\int\diff t\ \dot{x}_iV_i\ .
	\end{equation}
	This shows that, for arbitrary $\beta$,
	\begin{equation}
		(\partial_\theta+R^\beta_\theta)S^{\mathrm{b}}\=0\ .
	\end{equation}

	\noindent\textbf{Determinant matching.\ }
	Now we check the infinitesimal determinant-matching condition
	\begin{equation}\label{eq:det_match_cond}
		(\partial_g+R_g)S^{\mathrm{f}}[x]\=\im \int \diff t\ \diff t'\ \sfrac{\delta K_i[x,t]}{\delta x_j(t')}\;\;\delta_{ij}\delta(t-t')\ ,
	\end{equation}
	with
	\begin{equation}
		R_g^\pm\=\int\diff t\ K_i^\pm[x,t]\sfrac{\delta}{\delta x_i(t)}\ ,\qquad K_i^+[x,t]\=\int\diff \tau\ \partial_g V_k(\tau)S_{ki}(t,\tau)\ ,\quad K_i^-[x,t]\=\int\diff \tau\ \partial_g V_k(\tau)S_{ki}(\tau,t)
	\end{equation}
	and the nonlocal part of the action,
	\begin{equation}
		S^{\mathrm{f}}[x]\=-\im\ \ln\ \det \bigl[-\im(\im\sfrac{\diff}{\diff t}\delta_{ij}-V_{ij})\delta(t-t')\bigr]\=-\im\ \int\diff t\ \diff t'\ \ln\bigl[-\im(\im\sfrac{\diff}{\diff t}\delta_{ij}-V_{ij})\delta(t-t')\bigr]\ \delta_{ij}\;\delta(t-t')\ .
	\end{equation}
	With
	\begin{equation}
		\sfrac{\delta S_{ki}(\tau,t)}{\delta x_j(t')}\=S_{kl}(\tau,t')V_{ljm}(t')S_{mi}(t',t) \quad\und\quad
		\sfrac{\delta S_{ki}(t,\tau)}{\delta x_j(t')}\=S_{kl}(t',\tau)V_{ljm}(t')S_{mi}(t,t')
	\end{equation}
	we find on the right-hand side of \eqref{eq:det_match_cond} that
	\begin{equation}
		\begin{aligned}
			\im\int \diff t\ \diff t'\ \sfrac{\delta K_i[x,t]}{\delta x_j(t')}\;\;\delta_{ij}\delta(t-t')\=&\im\int\diff t\ \partial_g V_{ij}(t)S_{ji}(t,t)\\
			\ +\ &\im\int\diff t\ \diff t'\ \partial_g V_k(t)[S_{kl}(t',t)+S_{kl}(t,t')]V_{lim}(t')S_{mi}(t',t')\\+\ \theta\ 
			&\im\int\diff t\ \diff t'\ \partial_g V_k(t)[S_{kl}(t',t)-S_{kl}(t,t')]V_{lim}(t')S_{mi}(t',t')\ .
		\end{aligned}
	\end{equation}
	Using the chain rule it is easy to check that this indeed matches the left hand side of \eqref{eq:det_match_cond}.
	
	We can perform the analogous check for the theta flow,
	\begin{equation}\label{eq:det_match_cond_theta}
		(\partial_\theta+R^\beta_\theta)S^{\mathrm{f}}[x]\=\im\int \diff t\ \diff t'\ \sfrac{\delta K^\theta_i[x,t]}{\delta x_j(t')}\;\;\delta_{ij}\delta(t-t')\ ,
	\end{equation}
	where $\partial_\theta S^{\mathrm{f}}[x]=0$ and
	\begin{equation}
		K^\theta_i[x,t]\=\sfrac{\beta}{2} \int\diff \tau\ V_k(\tau)[S_{ki}(t,\tau)-S_{ki}(\tau,t)]\ +\ \im\  \sfrac{\beta-1}{2}\int\diff \tau\ \dot{x}_k(\tau)[S_{ki}(t,\tau)+S_{ki}(\tau,t)]\ .
	\end{equation}
	In a calculation similar to the one for the $g$~flow, we find
	\begin{equation}
		\begin{aligned}
			\im\int \diff t\ \diff t'\ \sfrac{\delta K^\theta_i[x,t]}{\delta x_j(t')}\;\;\delta_{ij}\delta(t-t')
			\=
			\im\ \sfrac{\beta}{2}&\int\diff t\ \diff t'\  V_k(t)[S_{kl}(t',t)-S_{kl}(t,t')]V_{lim}(t')S_{mi}(t',t')\\
			-\ \sfrac{\beta-1}{2}&
			\int\diff t\ \diff t'\  \dot{x}_k(t)[S_{kl}(t',t)+S_{kl}(t,t')]V_{lim}(t')S_{mi}(t',t')
			\ .
		\end{aligned}
	\end{equation}
 	It is again straightforward to see that this matches the left-hand side of \eqref{eq:det_match_cond_theta}.

	\section{Amplitudes from conventional Feynman perturbation theory}\label{app:feyn}
	After Wick rotating to Euclidean space, one finds the following Feynman rules in frequency space.
	\begin{itemize}
		\item The free fermion propagator is
		\begin{equation}
			\begin{tikzpicture}[baseline={([yshift=-.1cm]current bounding box.center)}]
				\begin{feynman}
					\vertex (a);
					\vertex [right=of a] (b);
					\diagram* {
						(a) -- [fermion] (b),
					};
				\end{feynman}\;\end{tikzpicture}\quad \=\quad\frac{\im\omega-m}{\omega^2+m^2}\ .
		\end{equation}
		\item The free boson propagator is
		\begin{equation}
			\begin{tikzpicture}[baseline={([yshift=-.1cm]current bounding box.center)}]
				\begin{feynman}
					\vertex (a);
					\vertex [right=of a] (b);
					\diagram* {
						(a) -- [photon] (b),
					};
				\end{feynman}\;\end{tikzpicture}\quad \=\quad\frac{-1}{\omega^2+m^2}\ .
		\end{equation}
		\item The vertices are
		\begin{equation}
			\begin{tikzpicture}[baseline={([yshift=-.1cm]current bounding box.center)},scale=0.5, transform shape]
			\begin{feynman}
						\vertex (a);
						\vertex [right=of a] (b);
						\vertex [above left=of a] (a1);
						\vertex [below left=of a] (a2);
						\diagram*{
							(a) -- [photon] (b),
							(a1) -- [photon] (a),
							(a2) -- [photon] (a),
						};\end{feynman}
			\end{tikzpicture}\quad \=\quad 3!\ mg\ ,\qquad\quad
			\begin{tikzpicture}[baseline={([yshift=-.1cm]current bounding box.center)},scale=0.5, transform shape]
			\begin{feynman}
						\vertex (a);
						\vertex [right=of a] (a1);
						\vertex [left=of a] (a2);
						\vertex [below=of a] (a3);
						\vertex [above=of a] (a4);
						\diagram*{
							(a1) -- [photon] (a) -- [photon] (a2),
							(a3) -- [photon] (a) -- [photon] (a4),
						};\end{feynman}
			\end{tikzpicture}\quad \=\quad \sfrac{4!}{2}\ g^2\ ,\qquad\quad
			\begin{tikzpicture}[baseline={([yshift=-.1cm]current bounding box.center)},scale=0.5, transform shape]
					\begin{feynman}
						\vertex (a);
						\vertex [right=of a] (b);
						\vertex [above left=of a] (a1);
						\vertex [below left=of a] (a2);
						\diagram*{
							(a) -- [photon] (b),
							(a1) -- [fermion] (a),
							(a2) -- [anti fermion] (a),
						};\end{feynman}
			\end{tikzpicture}\quad \=\quad 2\ g\ .
		\end{equation}
		\item We need to enforce momentum conservation at every vertex. For each fermion loop, one has to include a factor of $-1$. 
		Each loop frequency $l$ comes with an integral $\int\sfrac{\diff \omega}{2\pi}$. 
		Lastly, one has to divide by the symmetry factor of the diagram, i.e.~the number of permutations of internal lines that leave the diagram invariant.
		\item We use the convention where all external frequencies are outgoing.
	\end{itemize}

	\noindent\textbf{One-point function.\ }
	At order $g$ there are two connected diagrams contributing to the bosonic one-point function. Pulling out the prefactor
	\begin{equation}
		-g\sfrac{2\pi\delta(\omega)}{m^2}\ ,
	\end{equation}
	they are
	\begin{equation}
		\begin{tikzpicture}[baseline=(current bounding box.center)]
			\begin{feynman}
				\vertex (s) at ($(b) + (0, .7cm)$);
				\draw[photon]  (b) -- (s);
				\draw (s) [photon] arc [start angle=-90, end angle=270, radius=0.5cm];
			\end{feynman}
		\end{tikzpicture}
		\quad\rightarrow\quad \sfrac{3!}{2}m\int\sfrac{\diff l}{2\pi}\sfrac{-1}{l^2+m^2}\=-\sfrac{3}{2}\ ,
	\end{equation}
	\begin{equation}
		\begin{tikzpicture}[baseline=(current bounding box.center)]
			\begin{feynman}
				\vertex (s) at ($(b) + (0, .7cm)$);
				\draw[photon]  (b) -- (s);
				\draw[fermion] (s) arc [start angle=-90, end angle=270, radius=0.5cm];
			\end{feynman}
		\end{tikzpicture}
		\quad\rightarrow\quad -2\int\sfrac{\diff l}{2\pi}\sfrac{-m}{l^2+m^2}\=1\ ,
	\end{equation}
	so that
	\begin{equation}\label{eq:feyn_1point}
		\bigl\langle\x(\omega)\bigr\rangle_g\=g\sfrac{\pi\delta(\omega)}{m^2}\ +\ \mathcal{O}(g^3)\ .
	\end{equation}
	
	\noindent\textbf{Two-point function.\ }
	At order $g^2$ there are five connected diagrams contributing to the bosonic two-point function. Pulling out the prefactor
	\begin{equation}
		g^2\sfrac{2\pi\delta(\omega+\omega')}{(\omega^2+m^2)^2}\ ,
	\end{equation}
	there are three 1PI contributions,
	\begin{equation}
		\begin{tikzpicture}[baseline={([yshift=-.1cm]current bounding box.center)},scale=0.7, transform shape]
				\begin{feynman}
					\vertex (a);
					\vertex [right=of a] (b);
					\vertex [right=of b] (c);
					\vertex [right=of c] (d);
					\diagram*{
						(a) -- [photon] (b),
						(b) -- [fermion, half left, looseness=1.5] (c),
						(c) -- [fermion, half left, looseness=1.5] (b),
						(c) -- [photon] (d),
					};\end{feynman}
		\end{tikzpicture}\quad\rightarrow\quad 0\ ,
	\end{equation}
	\begin{equation}
		\begin{tikzpicture}[baseline={([yshift=-.1cm]current bounding box.center)},scale=0.7, transform shape]
			\begin{feynman}
				\vertex (a);
				\vertex [right=of a] (b);
				\vertex [right=of b] (c);
				\vertex [right=of c] (d);
				\diagram*{
					(a) -- [photon] (b),
					(b) -- [photon, half left, looseness=1.5] (c),
					(c) -- [photon, half left, looseness=1.5] (b),
					(c) -- [photon] (d),
				};\end{feynman}
		\end{tikzpicture}\quad\rightarrow\quad\sfrac{3!\ 3!}{2}m^2\int\sfrac{\diff l}{2\pi}\sfrac{1}{l^2+m^2}\sfrac{1}{(l+\omega)^2+m^2}\=\sfrac{18m^2}{m(4m^2+\omega^2)}\ ,
	\end{equation}
	\begin{equation}
		\qquad
		\begin{tikzpicture}[baseline={([yshift=-.5cm]current bounding box.center)},scale=0.9, transform shape]
			\begin{feynman}
				\vertex (a);
				\vertex [right=of a] (b);
				\vertex [right=of b] (c);
				\vertex [above =1cm of b] (d);
				\diagram* {
					(a) -- [photon] (b) -- [photon] (c),
					(b) -- [photon, half left, looseness=1.5] (d),
					(b) -- [photon, half right, looseness=1.5] (d),
				};
			\end{feynman}
		\end{tikzpicture}\quad\rightarrow\quad \sfrac{4!}{2\cdot 2}\int\sfrac{\diff l}{2\pi}\sfrac{-1}{l^2+m^2}\=-\sfrac{3}{m}\ ,
	\end{equation}
	and two 1PR contributions known from the one-point function,
	\begin{equation}
		\begin{tikzpicture}
			\begin{feynman}
				\diagram [layered layout, horizontal=b to c] {
					a -- [photon] b -- [photon]c,
				};
				\vertex (s) at ($(b) + (0, .7cm)$);
				\draw[photon]  (b) -- (s);
				\draw (s) [photon] arc [start angle=-90, end angle=270, radius=0.5cm];
			\end{feynman}
		\end{tikzpicture}
		\quad\rightarrow\quad \sfrac{3!\ 3!}{2}m^2\sfrac{-1}{m^2}\int\sfrac{\diff l}{2\pi}\sfrac{-1}{l^2+m^2}\=\sfrac{9}{m}\ ,
	\end{equation}
	\begin{equation}
		\begin{tikzpicture}
			\begin{feynman}
				\diagram [layered layout, horizontal=b to c] {
					a -- [photon] b -- [photon]c,
				};
				\vertex (s) at ($(b) + (0, .7cm)$);
				\draw[photon]  (b) -- (s);
				\draw[fermion] (s) arc [start angle=-90, end angle=270, radius=0.5cm];
			\end{feynman}
		\end{tikzpicture}
		\quad\rightarrow\quad -\sfrac{3!\ 2}{1}m\sfrac{-1}{m^2}\int\sfrac{\diff l}{2\pi}\sfrac{-m}{l^2+m^2}\=-\sfrac{6}{m}\ .
	\end{equation}
	Adding all the contributions, we obtain
	\begin{equation}\label{eq:feyn_2point}
		\bigl\langle\x(\omega)\x(\omega')\bigr\rangle_g\=2\pi\delta(\omega+\omega')G_0(\omega)\ +\ g^2\ \sfrac{2\pi\delta(\omega+\omega')}{(m^2+\omega^2)^2}\sfrac{18m^2}{4m^3+m\omega^2}\ +\ \mathcal{O}(g^3)\ .
	\end{equation}
	
	\noindent\textbf{Three-point function.\ }
	At order $g^3$ there are three 1PI connected diagrams contributing to the bosonic three-point function. The amputated diagrams are
	\begin{equation}
		\begin{aligned}
			F_1\=\begin{tikzpicture}[baseline={([yshift=-.1cm]current bounding box.center)},scale=.8,transform shape]
				\begin{feynman}
					\vertex (a1) at ($(0,0)$);
					\vertex (a2) at ($(1,0)$);
					\vertex (a3) at ($(.5,-.87)$);
					\vertex [above left=1cm of a1] (e1);
					\vertex [above right=1cm of a2] (e2);
					\vertex [below =1cm of a3] (e3);
					\diagram*{
						(a1) -- [photon] (a2) -- [photon] (a3) -- [photon] (a1),
						(a1) -- [photon] (e1),
						(a2) -- [photon] (e2),
						(a3) -- [photon] (e3),
					};\end{feynman}
			\end{tikzpicture}\quad \longrightarrow&\quad (3!m)^3\ \int_lG_0(l)G_0(l-\omega_3)G_0(l-\omega_3-\omega_2)\\
		&\=-\sfrac{216m^2[12m^2+\sfrac{1}{2}(\omega_1^2+\omega_2^2+\omega_3^2)]}{(4m^2+\omega_1^2)(4m^2+\omega_2^2)(4m^2+\omega_3^2)}\ ,
		\end{aligned}
	\end{equation}
	\begin{equation}
		F_2\=\begin{tikzpicture}[baseline={([yshift=-.1cm]current bounding box.center)},scale=.8,transform shape]
			\begin{feynman}
				\vertex (a1) at ($(0,0)$);
				\vertex (a2) at ($(1,0)$);
				\vertex (a3) at ($(.5,-.87)$);
				\vertex [above left=1cm of a1] (e1);
				\vertex [above right=1cm of a2] (e2);
				\vertex [below =1cm of a3] (e3);
				\diagram*{
					(a1) -- [fermion] (a2) -- [fermion] (a3) -- [fermion] (a1),
					(a1) -- [photon] (e1),
					(a2) -- [photon] (e2),
					(a3) -- [photon] (e3),
				};\end{feynman}
		\end{tikzpicture}\quad \longrightarrow\quad -2^3\int_lS_0(l)S_0(l-\omega_3)S_0(l-\omega_3-\omega_2)\=0\ ,
	\end{equation}
	\begin{equation}
		\begin{aligned}
			F_3\=\begin{tikzpicture}[baseline={([yshift=-.1cm]current bounding box.center)},scale=.8,transform shape]
				\begin{feynman}
					\vertex (b);
					\vertex [above left=of b](a1);
					\vertex [below left=of b](a2);
					\vertex [right =of b] (d);
					\vertex [right =of d] (e);
					\diagram*{
						(a1) -- [photon] (b),
						(a2) -- [photon] (b),
						(b) -- [photon, half left, looseness=1.5] (d)
						-- [photon, half left, looseness=1.5] (b),
						(d) -- [photon] (e),
					};\end{feynman}
			\end{tikzpicture}\quad \longrightarrow\quad &m\sfrac{1}{2}\sfrac{4!}{2}3!\int_l G_0(l)G_0(l-\omega_3)\ +\ (\omega_3\rightarrow\omega_2)\ +\ (\omega_3\rightarrow\omega_1)\\
			\=&36\ \sfrac{48m^4+8m^2(\omega_1^2+\omega_2^2+\omega_3^2)+\omega_1^2\omega_2^2+\omega_1^2\omega_3^2+\omega_2^2\omega_3^2}{(4m^2+\omega_1^2)(4m^2+\omega_2^2)(4m^2+\omega_3^2)}\ .
		\end{aligned}
	\end{equation}
	The remaining 1PR contributions can be taken from the results of the two- and three-point functions,
	\begin{equation}
		F_4\=\begin{tikzpicture}[baseline={([yshift=-.1cm]current bounding box.center)},scale=.8, transform shape]
			\begin{feynman}
				\vertex (b);
				\vertex [above left=of b](a1);
				\vertex [below left=of b](a2);
				\vertex [right =1cm of b] (c) [blob,label=\(\textrm{2-point}\)] {};
				\vertex [right =of c] (d);
				\diagram*{
					(a1) -- [photon] (b),
					(a2) -- [photon] (b),
					(b) -- [photon] (c) -- [photon] (d),
				};\end{feynman}
		\end{tikzpicture}\quad\longrightarrow\quad 3!m\sfrac{-18m^2}{(4m^3+m\omega_3^2)(m^2+\omega_3^2)}\ +\ (\omega_3\rightarrow\omega_2)\ +\ (\omega_3\rightarrow\omega_1)\ ,
	\end{equation}
	\begin{equation}
		F_5\=\begin{tikzpicture}[baseline={([yshift=-.1cm]current bounding box.center)},scale=.8, transform shape]
			\begin{feynman}
				\vertex[blob,label=\(\textrm{1-point}\)] (blob) at (1, 0) {};
				\vertex (a) at (-1,0);
				\vertex (b) at (0,1);
				\vertex (c) at (0,-1);
				\vertex (o) at (0,0);
				\diagram* {
					(a) -- [photon] (o) -- [photon] (blob),
					(b) -- [photon] (o),
					(c) -- [photon] (o),
				};
			\end{feynman}
		\end{tikzpicture}\quad\longrightarrow\quad \sfrac{6}{m^2}\ .
	\end{equation}
	We write down the contributions in the same way as we did for the Nicolai calculation, 
	pulling out the prefactor~\eqref{eq:prefactor} and writing the remaining contributions in terms of the symmetric polynomials $t_2$ and $t_3$~\eqref{eq:sym_pol}:
	\begin{equation}
		\begin{aligned}
			&F_1\quad\rightarrow\quad 2592 m^{10} - 5400 m^8 t_2 + 3024 m^6 t_2^2 - 216 m^4 t_2^3 + 
			2592 m^4 t_3^2 - 216 m^2 t_2 t_3^2\ ,\\
			&F_2\quad\rightarrow\quad 0\ ,\\
			&F_3\quad\rightarrow\quad -1728 m^{10} + 4032 m^8 t_2 - 2916 m^6 t_2^2 + 648 m^4 t_2^3 - 
			36 m^2 t_2^4 - 1728 m^4 t_3^2 + 576 m^2 t_2 t_3^2 - 36 t_2^2 t_3^2\ ,\\
			&F_4\quad\rightarrow\quad 5184 m^{10} - 8640 m^8 t_2 + 4428 m^6 t_2^2 - 1080 m^4 t_2^3 + 
			108 m^2 t_2^4 - 1620 m^4 t_3^2 + 432 m^2 t_2 t_3^2\ ,\\
			&F_5\quad\rightarrow\quad -384 m^{10} + 960 m^8 t_2 - 792 m^6 t_2^2 + 240 m^4 t_2^3 - 24 m^2 t_2^4 - 
			390 m^4 t_3^2 + 204 m^2 t_2 t_3^2 - 30 t_2^2 t_3^2 - 6 m^{-2} t_3^4\ .
		\end{aligned}
	\end{equation}
	The 1PI and 1PI sums become
	\begin{equation}
		\begin{aligned}
			&F_{\mathrm{1PI}}\=F_1+F_2+F_3\\ &\rightarrow\quad 864 m^{10} - 1368 m^8 t_2 + 108 m^6 t_2^2 + 432 m^4 t_2^3 - 36 m^2 t_2^4 + 
			864 m^4 t_3^2 + 360 m^2 t_2 t_3^2 - 36 t_2^2 t_3^2\ ,
		\end{aligned}
	\end{equation}
	\begin{equation}
		\begin{aligned}
			&F_{\mathrm{1PR}}\=F_4+F_5\\ &\rightarrow\quad 4800 m^{10} - 7680 m^8 t_2 + 3636 m^6 t_2^2 - 840 m^4 t_2^3 + 
			84 m^2 t_2^4 - 2010 m^4 t_3^2 + 636 m^2 t_2 t_3^2 - 30 t_2^2 t_3^2 - 
			6 m^{-2} t_3^4\ ,
		\end{aligned}
	\end{equation}
	respectively. The final result (modulo the prefactor) reads
	\begin{equation}\label{eq:feyn_3point}
		\begin{aligned}
			&F\=F_{\mathrm{1PI}}+F_{\mathrm{1PR}}\\ &\rightarrow\quad 5664 m^{10} - 9048 m^8 t_2 + 3744 m^6 t_2^2 - 408 m^4 t_2^3 + 
			48 m^2 t_2^4 - 1146 m^4 t_3^2 + 996 m^2 t_2 t_3^2 - 66 t_2^2 t_3^2 - 
			6 m^{-2} t_3^4\ .
		\end{aligned}
	\end{equation}
	This exactly matches the result from the Nicolai computation~\eqref{eq:N_final}.

\end{document}